\documentclass[12pt]{article}

\usepackage{graphicx}
\usepackage{dcolumn}
\usepackage{bm}
\usepackage{soul}
\usepackage{ulem}
\usepackage{color}
\usepackage[T1]{fontenc}
\usepackage[super]{nth}
\usepackage{amsmath}
\usepackage[center]{subfigure}
\usepackage[utf8]{inputenc}
\usepackage{array}
\usepackage{tabu}
\usepackage[super]{nth}
\usepackage{tikz}
\usepackage{standalone}
\usepackage{amssymb}
\usepackage{cite}
\usepackage{float}
\usepackage[version=3]{mhchem}
\usepackage{pstricks}
\usepackage{braket}
\usetikzlibrary{snakes}
\usepackage{picture}

\def\beq{\begin{equation}}
\def\eeq{\end{equation}}
\def\ba{\begin{eqnarray}}
\def\ea{\end{eqnarray}}

\begin{document}

\begin{center}
{\Large{\bf Quantum Chaotic Systems and Random Matrix Theory}} \\
\ \\
\ \\
by \\
Akhilesh Pandey, Avanish Kumar and Sanjay Puri \\
School of Physical Sciences, Jawaharlal Nehru University, New Delhi -- 110067, India.
\end{center}

\begin{abstract}
This article is an introductory review of random matrix theory (RMT) and its applications, with special focus on quantum chaos.
Random matrices were first used by Wigner to understand the spectra of complex nuclei from a statistical perspective. Subsequently there have been novel applications to diverse areas, e.g., atomic and molecular physics, mesoscopic and nanoscopic systems, microwave cavities, econophysics, biological sciences, communication theory. This article is designed to be accessible at the graduate and post-doctoral level. {\color{red} (To appear in 21st Century Nanoscience: A Handbook, edited by K.D. Sattler, CRC Press.)}
\end{abstract}

\section{Introduction}\label{S1}

This article provides a pedagogical introduction to the subject of random matrix theory (RMT) and its applications. More advanced readers may refer to several books and reviews which deal extensively with this subject \cite{CEP, WIGNER1, BRODY, BG, CWB, GMW, FH, ML, ALHS, HS, CH, CWBHB}.
Random matrices were introduced in the context of statistical multivariate analysis by Wishart \cite{WILKS}. They were first used in physics by Wigner to understand energy levels in complex nuclei \cite{ WIGNER1}. In Wigner's prescription, ensembles of random matrices were used to model Hamiltonians of complex systems. The mathematical development of RMT was done by Wigner himself, as well as Dyson \cite{FJDY1, FJDY2, FJDY3, FJDY4, FJDY5}, Mehta \cite{ML1}, Gaudin \cite{MLG}, Porter \cite{CEP} and several others. The ensemble properties of interest to physicists are energy level spectra, decay widths, and fluctuations of scattering cross-sections. In an important discovery \cite{BGS}, it was found that random matrices are applicable to quantum chaotic systems, i.e., quantum systems whose classical counterparts are chaotic. This gave a justification for the utility of RMT to understand complex systems. Interest has also focused on transmission properties of mesoscopic/nanoscopic systems, quantum field theory, number theory, and communication theory. More recently, there have been many studies of time series invoking the original ideas of Wishart \cite{WILKS}. These have found applications in diverse areas such as econophysics, atmospheric sciences, biology etc.

In physics applications, ensembles are classified according to the time-reversal and space-rotation symmetries of the underlying physical systems. The three important classes are Gaussian ensembles of symmetric Hermitian matrices, general complex Hermitian matrices, and quaternion self-dual Hermitian matrices. These are referred to as Gaussian orthogonal ensembles (GOE), Gaussian unitary ensembles (GUE), and Gaussian symplectic ensembles (GSE). (These are also known as \textit{classical ensembles}.)
Here the terms orthogonal, unitary, and symplectic refer to the transformations under which ensembles remain invariant.
Dyson \cite{FJDY1, FJDY2, FJDY3} introduced analogous ensembles involving unitary matrices. These are referred to as the \textit{circular ensembles}. They have the same three-fold classification as above, giving rise to COE, CUE, and CSE for symmetric unitary, general unitary and self-dual unitary matrices respectively.

This article is organized as follows. In Sec.~\ref{S2}, we present definitions of the ensembles. In Sec.~\ref{S3}, we discuss joint probability distributions (jpd) of the eigenvalues and their fluctuation measures. In Sec.~\ref{S4}, we discuss the statistical properties of eigenvectors. In Sec.~\ref{S5}, we elucidate the importance of these ensembles in analysis of nuclear spectra. In Sec.~\ref{S6},  we give a brief history of the connection between quantum chaos and RMT. In this context, we will use quantum kicked rotors (QKR) as a paradigm of quantum chaos. In Sec.~\ref{S7}, we highlight the connection between quantum counterparts of integrable systems and Poisson statistics. In Sec.~\ref{S8}, we introduce the classical kicked rotor and its properties. In Sec.~\ref{S9}, we define the QKR. In Sec.~\ref{S10}, we give results for eigenvalues and eigenvector fluctuations in QKR. In Sec.~\ref{S11}, we discuss transition ensembles, viz., ensembles which are intermediate between the classical ones. In Sec.~\ref{S12}, we turn our attention to mesoscopic systems. We discuss the experimentally important phenomenon of universal conductance fluctuations (UCF) and its understanding via RMT. In Sec.~\ref{S13}, we briefly discuss finite-range Coulomb gas (FRCG) models and their applications to QKR. In Sec.~\ref{S14}, we discuss the Wishart ensemble and its applications. Section~\ref{S15} concludes this paper.

\section{Random Matrix Theory}\label{S2}

In quantum systems we encounter two types of matrices: Hermitian matrices which represent Hamiltonians, and unitary matrices which represent scattering matrices and time-evolution operators. Dyson has shown \cite{FJDY5} that there are three important classes of matrices, depending on the time-reversal invariance (TRI), and space-rotation invariance (SRI). In Table 1, we list these different classes.
\begin{table*}
\centering
\caption{CLASSIFICATION OF RANDOM MATRICES}
\begin{tabular}{@{}ccccc@{}}
 \hline \\
~~TRI~~~& SRI ~~& $H$ or $U$~~~& $\beta$ ~~& Group \\
\\
\hline
Yes~~&  Yes ~~~~&  Symmetric       &  1  & ~~(O) \\
Yes~~&  No but integral spin~~~~&  Symmetric  &  1  & ~~(O) \\
Yes~~&  No but half-integral spin ~~~ &  Quaternion self dual  &  4  & ~~(S) \\
No~~&  Irrelevant  ~~~~~~ &  General         &  2  & ~~(U) \\
 \hline
\end{tabular}
\end{table*}
In this table, $H$ and $U$ refer to hermitian and unitary matrices. We also specify the transformation groups under which $H$ and $U$ remain in the same class \cite{CEP, WIGNER2}. The parameter $\beta$ (also known as the Dyson parameter) denotes the number of distinct components in the off-diagonal elements of $H$ or $U$.

For $\beta=1$, the matrices are symmetric in both $H$ and $U$ cases. For $\beta=4$, they are self-dual (see Appendix A). For $\beta=2$, there are no such constraints on matrices. For other symmetries, the matrices may be of block-diagonal form. For simplicity, we will mostly consider matrices with only a single block.
The jpd of the matrix elements of $A$, where the $N$-dimensional matrix $A$ may be of the type $H$ or $U$, is
\begin{equation}\label{Eq.1}
P(A)= C\exp(-\beta~tr~V(A)).
\end{equation}
Here, $V(A)$ is a positive-definite function of $A$, and $C$ is the normalization constant. (In this paper, $C$ and its variants will denote different normalization constants.) The cases studied extensively are (a) the Gaussian ensembles with $V(\xi)=\xi^{2}/4v^{2}$ for the Hermitian case (with $v$ setting the scale of matrix elements); and (b) the (uniform) circular ensembles with $V(\xi)=0$. These are commonly referred to by their acronyms GOE, GUE, and GSE in the Gaussian cases \cite{ML, WIGNER1, BRODY}, and COE, CUE, and CSE in the circular cases. The letters O (orthogonal), U (unitary), and S (symplectic) specify the group transformation invariances in Table 1.

The GOE matrices are symmetric, and the distinct matrix elements $A_{jk}$ have independent Gaussian distributions with zero mean and variance $(1+\delta_{jk})v^{2}$. For GUE, the matrix elements are complex. The real part has the same distribution as GOE, and the imaginary part is real anti-symmetric with zero mean and the variance $(1-\delta_{jk})v^{2}$. For GSE, one needs four matrices. One of these is symmetric as in the GOE, and three are anti-symmetric as described above for GUE. The construction of GSE matrices has been described in Appendix A.

To study the statistical properties of eigenvalues and eigenvectors, we transform the jpd in Eq.~(\ref{Eq.1}) to a jpd of eigenvalues and eigenvectors. The exponential factor transforms into a corresponding factor for eigenvalues alone as the trace of $V(A)$ is invariant under transformations of the relevant group. There is an additional Jacobian factor which decomposes into a product of eigenvalues and eigenvector-dependent functions. Thus the eigenvalue and eigenvector distributions are independent of each other.

For $\beta=4$, the eigenvalues are doubly degenerate. In this case we have $2N$-dimensional matrices and therefore $2N$ eigenvalues. However, in the discussion of statistical properties, it is standard practice to consider only the $N$ distinct eigenvalues. We will follow this convention in all subsequent sections except in Sec.~\ref{S11}, where the GSE $\rightarrow$ GUE transition requires explicit consideration of the double degeneracy.

\section{Statistical Properties of Eigenvalues}\label{S3}

The transformation from matrix element space to eigenvalue and eigenvector space leads to the jpd of eigenvalues $\{x_i\}$ as 
\begin{equation}\label{Eq.2}
p(x_{1},\cdots,x_{N})= C^{\prime}\exp(-\beta W),
\end{equation}
where
\begin{equation}\label{Eq.3}
W=-\sum_{j< k}\log|x_{j}-x_{k}|+\sum_{j}V(x_{j}).
\end{equation}
The resultant jpd is 
\begin{equation}\label{Eq.4}
p(x_{1}, \cdots, x_{N})= C^{\prime} \prod_{j<k}|x_{j}-x_{k}|^{\beta}\prod_{j=1}^{N} e^{-\beta V(x_{j})}.
\end{equation}
The logarithmic part of Eq.~(\ref{Eq.3}) is derived from the Jacobian of the transformation. This contribution is reminiscent of a two-dimensional Coulomb potential, and therefore the system is termed the Coulomb gas \cite{FJDY1}. We will also refer to these ensembles as linear ensembles.

We remark that the eigenvalues for circular ensembles are of the form $e^{i\theta_j}$, and their fluctuations will refer to fluctuations of the eigenangles $\theta_j$. Eqs.~(\ref{Eq.2})-(\ref{Eq.4}), with $x_{j}\rightarrow e^{i\theta_{j}}$, apply to the jpd of eigenangles for circular ensembles. We will use the term level to describe both $x$ and $\theta$.

In our subsequent discussion, we will consider large $N$, unless otherwise specified. Then, it can be shown that the fluctuation properties of eigenvalues are independent of $V$ for each $\beta$. This happens because $V$ provides only the scale of the local spectra, and the fluctuations are studied in terms of local average spacing. Further, $H$ and $U$ give the same fluctuations. Therefore, we will focus mostly on Gaussian linear and uniform circular ensembles here \cite{ML, BRODY}.

An important quantity is the average density of eigenvalues, which is defined as $\bar{\rho} (x) = \int \cdots \int dx_2 \cdots dx_N~p(x, x_{2}, \cdots, x_{N})$. In the Gaussian case, $\overline{\rho}(x)$ is given by the well-known Wigner's semicircle:
\begin{equation}\label{Eq.5}
\overline{\rho}(x)= \frac{2\sqrt{R^{2}-x^{2}}}{\pi R^{2}}, \quad R^{2}= 4\beta v^{2} N.
\end{equation}

\begin{figure}[H]
\begin{center}
\includegraphics[width=\linewidth, width=8cm]{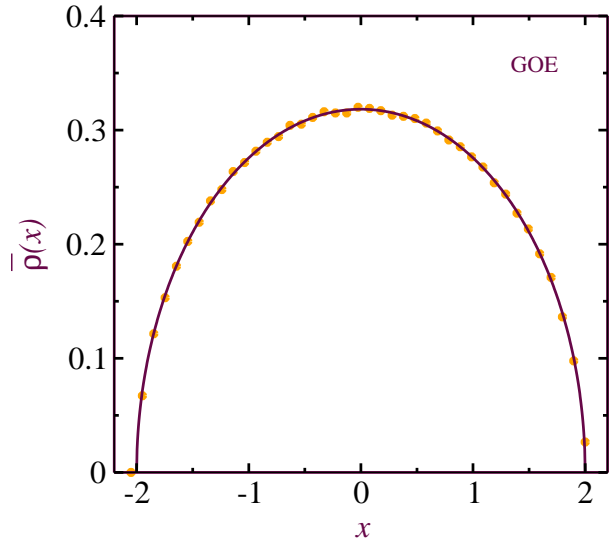}
\caption{Eigenvalue density of one GOE matrix for $N=5000$. The solid line corresponds to Wigner's semicircle in Eq.~(\ref{Eq.5}).}\label{fig:GOE}
\end{center}
\end{figure}

\begin{figure}[H]
\begin{center}
\includegraphics[width=\linewidth, width=6cm]{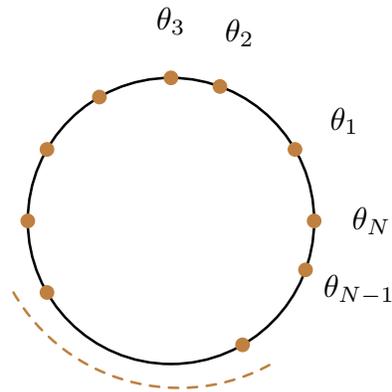}
\caption{A schematic diagram showing a set of $N$ eigenangles ($\theta_{1},\cdots, \theta_{N}$) in an ordered sequence on a circle. The eigenangles are treated as interacting particles which are restricted to move on the circle.}\label{fig:CIRCLE}
\end{center}
\end{figure}

Fig.~\ref{fig:GOE} shows the semicircular density for the GOE. In uniform circular ensembles, the density is
\begin{equation}\label{Eq.6}
\overline{\rho}(\theta)= \frac{1}{2\pi}, \quad 0 \leqslant \theta < 2\pi.
\end{equation}
In this case the average spacing for eigenvalues is $2\pi/N$ (Fig.~\ref{fig:CIRCLE}). Thus one needs to study statistical properties of $\xi_{j}=\theta_{j}N/(2\pi)$, so that the average spacing is unity everywhere.

\begin{figure}
\begin{center}
\includegraphics[width=\linewidth, width=10cm]{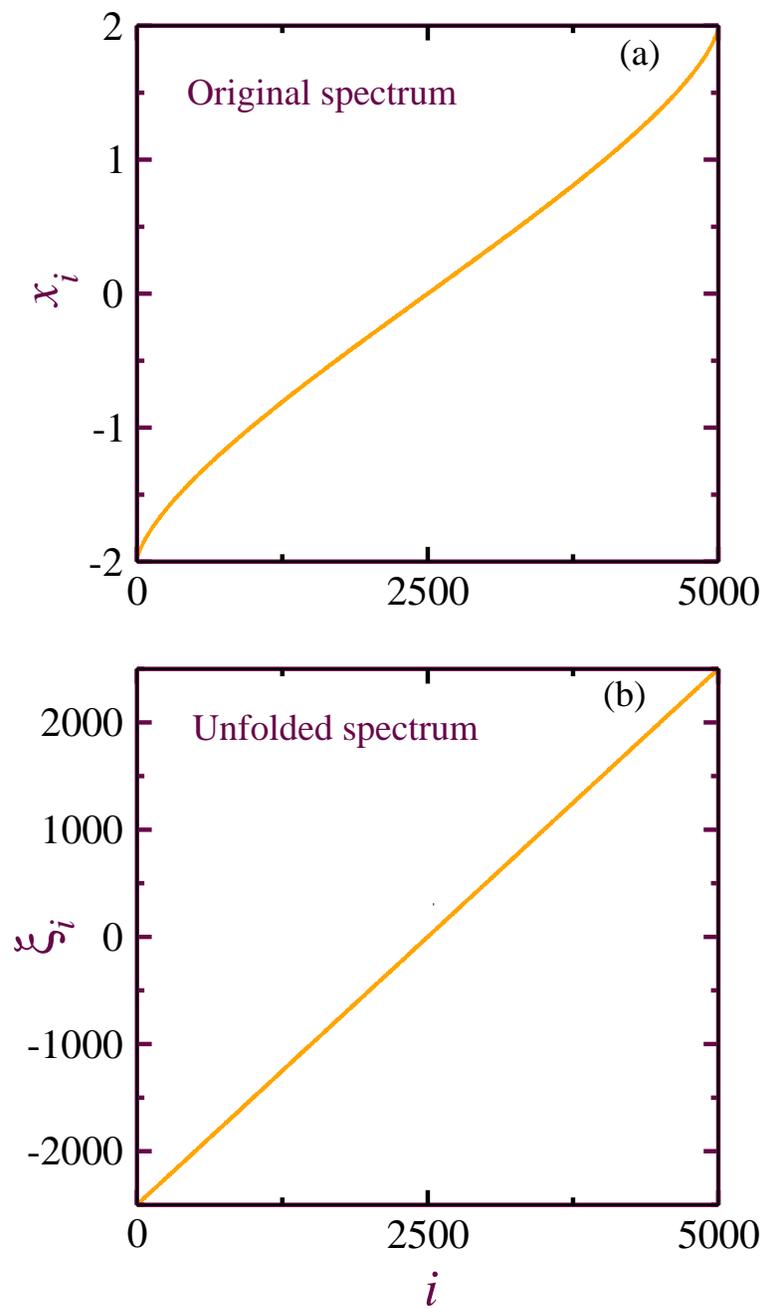}
\caption{Plot of (a) original and (b) unfolded spectrum of one GOE matrix with $N=5000$.}\label{fig:UNFOLDING}
\end{center}
\end{figure}

In many real systems the density is non-uniform (as in Eq.~(\ref{Eq.5})). In that case, one needs to `unfold' the spectrum so that the average spacing becomes 1 in the entire spectrum. The unfolding function is defined by
\begin{equation}\label{Eq.7}
F(\zeta)= N\int^{\zeta}\overline{\rho}(\zeta^{\prime})d\zeta^{\prime},
\end{equation}
where $\zeta^{\prime}$ refers to $x$ or $\theta$ depending on the system. The unfolding function for the semicircular density of Eq.~(\ref{Eq.5}) is 
\begin{equation}\label{Eq.8}
F(x)=\frac{N}{\pi R^{2}}\left[  x \sqrt{R^{2}-x^{2}} +R^{2}\arcsin\left( \frac{x}{R} \right)  \right]. 
\end{equation}
The unfolded spectrum, $\xi_{j}$, is now given by $\xi_{j}=F(x_{j})$ for the Gaussian ensembles (see Fig.~\ref{fig:UNFOLDING}).
%%%%%%%
When the Coulomb interaction (i.e., the logarithmic part in Eq.~(\ref{Eq.3})) is absent, then the jpd can be written as a product of positive-definite functions $w(x)$:
\begin{equation}\label{Eq.9}
p(x_{1},\cdots,x_{N})= C^{\prime}\prod_{j=1}^N w(x_{j}),
\end{equation}
where
\begin{equation}\label{Eq.10}
w(x)= \exp(-\beta V(x)).
\end{equation}
In this case, $\bar{\rho}(x)=w(x)$, and one can use Eq.~(\ref{Eq.7}) for unfolding the eigenvalue spectra. The jpd of the unfolded eigenvalues $\xi_{j}$ will be simply 1, with $ 0\leqslant \xi_{j} \leqslant N$. Thus $\xi_{j}$ will be distributed independently with uniform probability in $[0, N]$. This is referred to as the \textit{Poisson ensemble}.

\begin{figure}[H]
\begin{center}
\includegraphics[width=\linewidth, width=14cm]{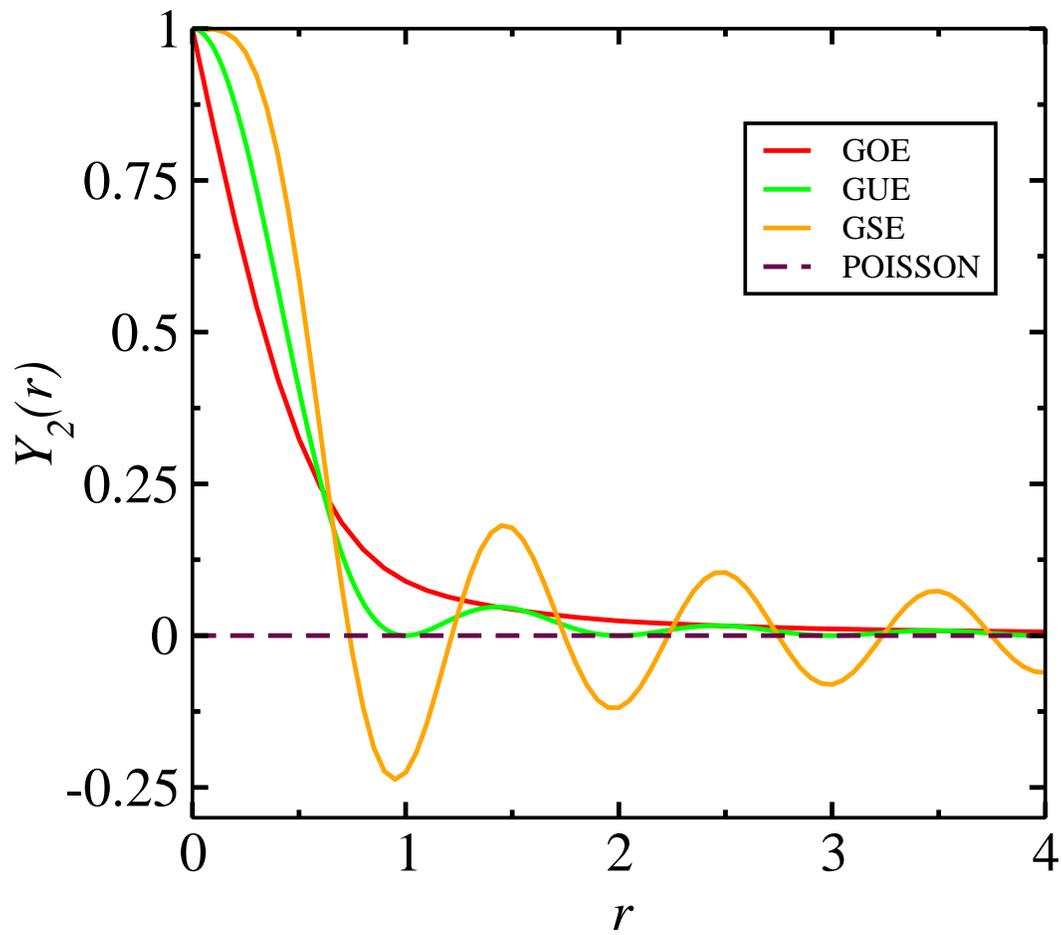}
\caption{Two-point cluster function for the classical ensembles corresponding to $\beta= 1, 2, 4$. The dashed line corresponds to The Poisson ensemble.}\label{fig:Y2S}
\end{center}
\end{figure}

The eigenvalue fluctuations are studied in terms of the correlation functions. The most important quantity is the two-level correlation function $R_{2}(r)$ or, equivalently, the two-level cluster function $Y_{2}(r)=1-R_{2}(r)$.
$R_{2}(r)dr$ is the conditional probability of finding a level in the interval $[\xi_{0}+r, ~\xi_{0}+r+dr]$, given that there is an eigenvalue at $\xi_{0}$. For the above three classes of ensembles, the (universal) cluster function is given by
\begin{align}
Y_{2}(r)&=\left(s(r)\right)^{2}+ \left(\frac{d}{dr}s(r)\right)\left(\int_{r}^{\infty}s(t)dt\right), \quad \beta=1, \label{Eq.11} \\
Y_{2}(r)&= \left(s(r)\right)^{2}, \quad \beta=2, \label{Eq.12}	\\
Y_{2}(r)&=\left(s(2r)\right)^{2}-\left(\frac{d}{dr}s(2r)\right)\left(\int_{0}^{r}s(2t)dt\right), \quad \beta=4, \label{Eq.13}
\end{align}
where
\begin{equation}\label{Eq.14}
s(r)= \frac{\sin(\pi r)}{\pi r}.
\end{equation}
For $r \gtrsim 1$, $Y_{2}(r)$ falls off as $1/(\beta \pi r^2)$ after the oscillatory terms have been averaged out. One can also define the $n$-level cluster function, and the exact results are known for all these ensembles \cite{PG, GP}. We remark that the correlation functions satisfy the property of stationarity, i.e., independent of $\xi_{0}$, and ergodicity, i.e., spectral and ensemble averages are equal \cite{AP3}. For the Poisson ensemble, i.e., ensemble of independent eigenvalues, $Y_{2}(r)=0$. See Fig.~\ref{fig:Y2S} for plots of $Y_{2}(r)$ vs. $r$ for all four ensembles.

\begin{figure}[H]
\begin{center}
\includegraphics[width=\linewidth, width=14cm]{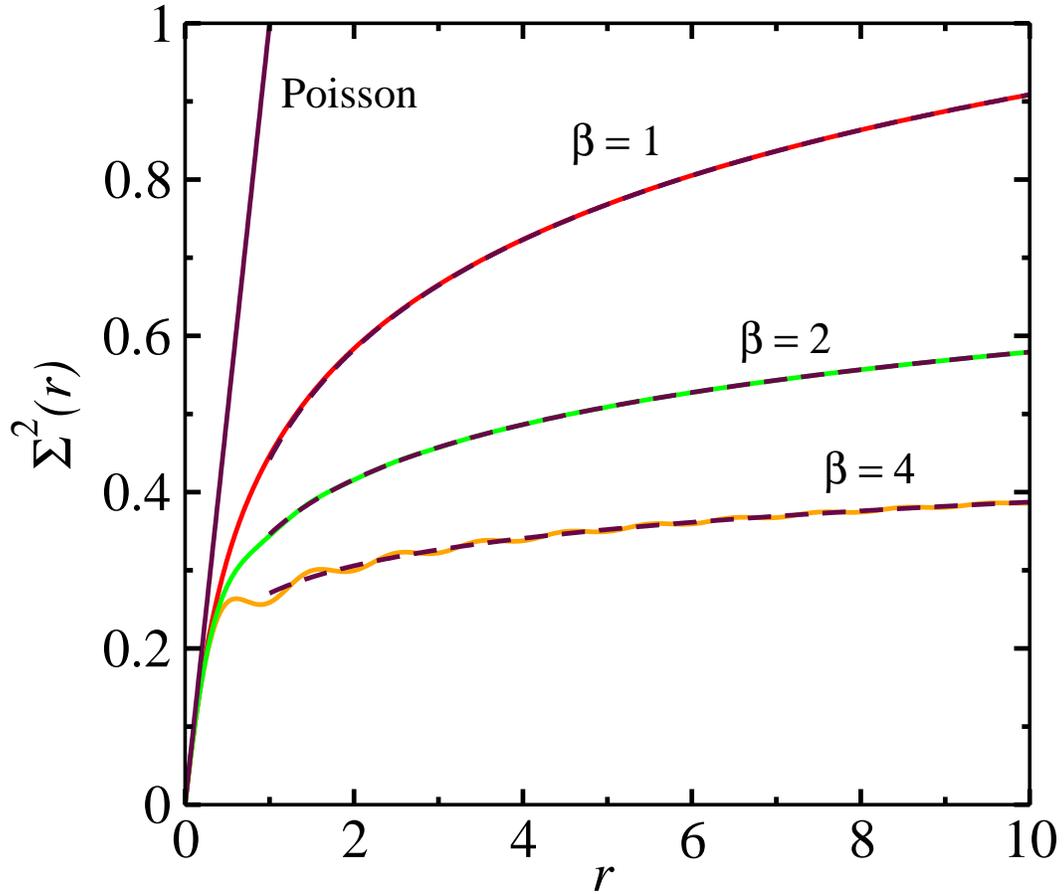}
\caption{Plot of number variance, $\Sigma^{2}(r)$ vs. $r$. The solid lines, corresponding to $\beta= 1, 2, 4$, have been calculated from Eqs.~(\ref{Eq.17})-(\ref{Eq.19}). The dashed lines correspond to Eq.~(\ref{Eq.16}), which is valid for $r\gtrsim 1$. For reference, the Poisson case has also been plotted. For small $r$ ($\lesssim 1$), $\Sigma^{2}(r)$ for classical ensembles follows linear behavior as in the Poisson case, but for large $r$ it shows logarithmic behavior.}\label{fig:NV}
\end{center}
\end{figure}

In actual applications, one usually considers the number variance $\Sigma^{2}(r)$, viz., the variance of the number of levels in intervals of length $r$. In terms of $Y_{2}$, it is given by
\begin{equation}\label{Eq.15}
\Sigma^{2}(r)= r-2\int_{0}^{r}(r-s)Y_{2}(s)ds.
\end{equation}
For the above ensembles,
\begin{equation}\label{Eq.16}
\Sigma^{2}(r)= \frac{2}{\beta \pi^{2}}\ln r + C_{\beta},
\end{equation}
valid for $r\gtrsim 1$. Here the constant $C_{\beta} = 0.4420, 0.3460, 0.2706$ for $\beta=1,2,4$ respectively.
The exact expressions for number variances are \cite{BRODY}
\begin{align}
\Sigma_{\beta=2}^{2}(r)= &\frac{1}{\pi^{2}}\left[\ln\left(2\pi r \right)+\gamma+1-\cos(2\pi r)-\text{Ci}\left(2\pi r\right) \right]\nonumber \\
& {} +r\left[ 1- \frac{2 \text{Si}\left( 2\pi r\right) }{\pi}\right]\label{Eq.17}, \\
\Sigma_{\beta=1}^{2}(r)= &2\Sigma_{\beta=2}^{2}(r)+ \left(\frac{\text{Si}(\pi r)}{\pi}\right)^{2} -\frac{\text{Si}(\pi r)}{\pi}\label{Eq.18}, \\
\Sigma_{\beta=4}^{2}(r)= &\frac{1}{2}\Sigma_{\beta=2}^{2}(2r)+ \left(\frac{\text{Si}(2\pi r)}{2\pi}\right)^{2} . \label{Eq.19} 
\end{align}
Here, $\gamma~(=0.5772\cdots)$ is the Euler constant and Si and Ci are sine and cosine integrals respectively. In contrast, for the Poisson ensemble, $\Sigma^{2}(r)=r$. (See Fig.~\ref{fig:NV}.)

\begin{figure}[H]
\begin{center}
\includegraphics[width=\linewidth, width=14cm]{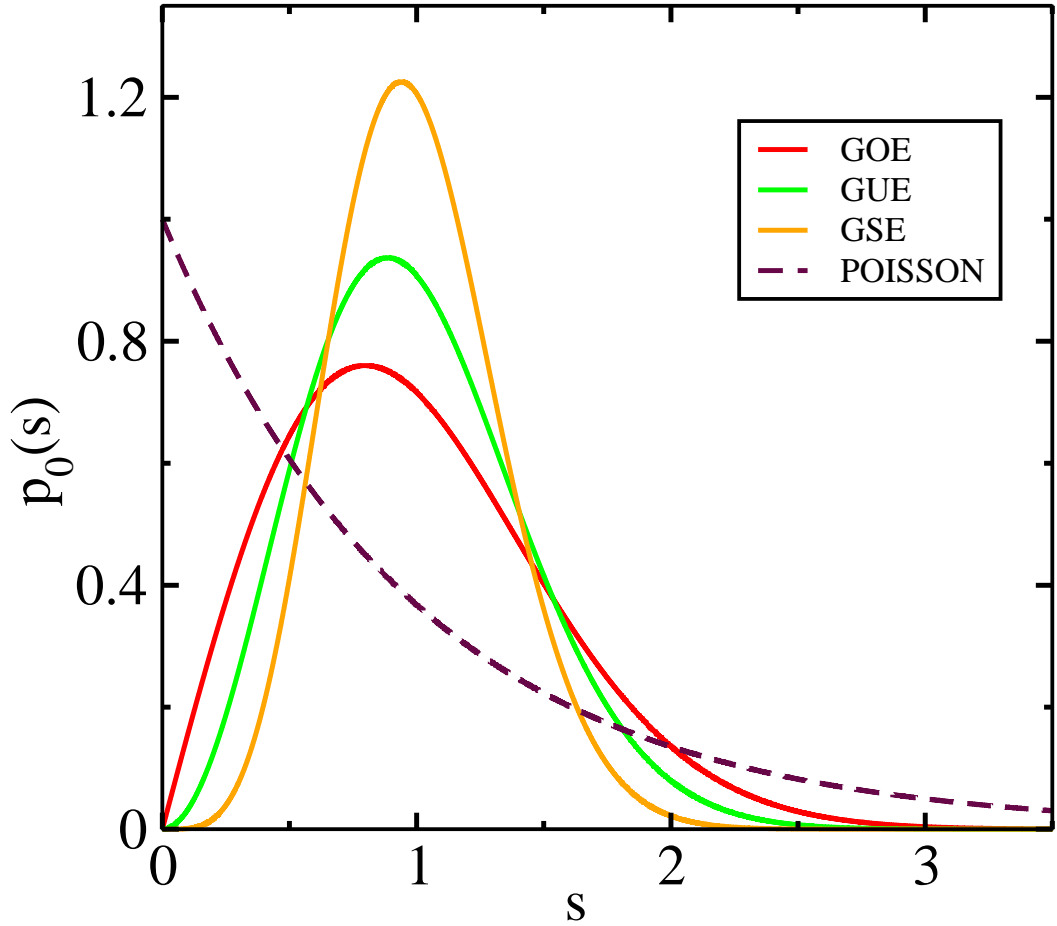}
\caption{Nearest-neighbor spacing distributions, $p_0(s)$ vs. $s$, from Eqs.~(\ref{Eq.20})-(\ref{Eq.22}) for classical ensembles. The dashed line corresponds to the Poisson ensemble. For small $s$, $p_{0}(s)\sim s^{\beta}$, whereas for large $s$, $p_{0}(s)\sim \exp(-s^{2})$ for $\beta = 1,2,4$.}\label{fig:NNSD}
\end{center}
\end{figure}

Another widely used fluctuation measure is the nearest-neighbor spacing distribution $p_{0}(s)$, where $s$ is the spacing between two consecutive (unfolded) eigenvalues, i.e., $s_{i}=\xi_{i+1}-\xi_{i}$. For the Poisson ensemble, $p_{0}(s)=e^{-s}$. For the three random matrix ensembles, the exact results for $p_{0}(s)$ have been derived by Mehta. These expressions are complicated but the spacing distributions for two-dimensional matrices gives excellent approximations to the exact results. These are (see Fig.~\ref{fig:NNSD}):
\begin{align}
p_{0}(s)&=\frac{\pi}{2}s\exp\left(-\frac{\pi}{4}s^{2}\right), &	\beta=1,  	\label{Eq.20} \\
p_{0}(s)&=\frac{32}{\pi^{2}}s^{2}\exp\left(-\frac{4}{\pi}s^{2}\right), &	\beta=2,	 \label{Eq.21} \\
p_{0}(s)&=\frac{2^{18}}{3^{6}\pi^{3}}s^{4}\exp\left(-\frac{64}{9\pi}s^{2}\right),	& \beta=4.	\label{Eq.22}  
\end{align}

The higher-order spacing distributions $p_{k}(s)$ have also been studied. These are distributions of spacings between two eigenvalues with $k$ intermediate eigenvalues, i.e., $s_{i}=\xi_{i+k+1}-\xi_{i}$. For the nearest-neighbor spacing distribution, $k=0$. These functions satisfy
\begin{equation}\label{Eq.23}
R_{2}(s)= \sum_{k=0}^{\infty}p_{k}(s).
\end{equation}
Another quantity of interest is the spacing variance $\sigma^{2}(k)$, which is the variance of $p_{k}(s)$. This is closely related to $\Sigma^{2}(k+1)$. These two quantities are equal for the Poisson ensemble. On the other hand, for the three classical ensembles, the difference $\Sigma^{2}(k+1)-\sigma^{2}(k)$ is close to $1/6$, e.g., $0.161$ for $k=0$, and approaches $1/6$ rapidly as $k$ increases \cite{BRODY}.

\begin{figure}
\begin{center}
\includegraphics[width=\linewidth, width=12cm]{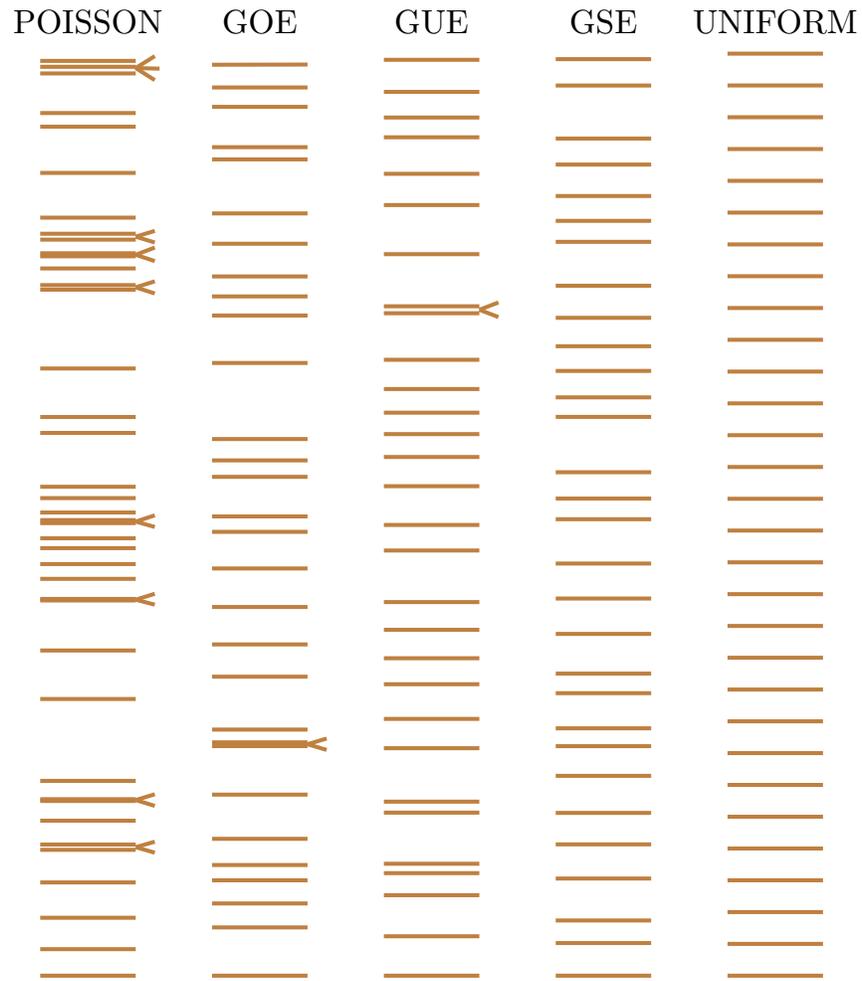}
\caption{Comparison of unfolded spectrum for different ensembles. We note that the degree of uniformity increases from the Poisson to GSE ensembles. The arrowheads mark the occurrences of pairs of levels with the spacings smaller than $1/2$ of the average spacing.}\label{fig:COMPARISON}
\end{center}
\end{figure}

To illustrate the difference between random matrix ensembles and the Poisson case, we consider a sequence of, say, 10,000 levels. A segment of this is shown in Fig.~\ref{fig:COMPARISON}.
%%%%%%%%%%%%%
In that case, $\Sigma(r) \simeq 1$ in the random matrix ensembles, whereas $\Sigma(r) \simeq 100$ for the Poisson case. This property of level correlations is referred to as the spectral rigidity, and is a consequence of the long-range correlations in the random matrix spectra. In descending order of rigidity, the spectra can be ramked as GSE, GUE and GOE. Further we mention that $p_{0}(s)$ and $R_{2}(s)$ approach 0 as $s^{\beta}$ for small $s$. This is referred to as level repulsion in random matrix ensembles. This should be contrasted with level clustering observed in Poisson ensembles, where $p_{0}(s)=1$ for $s \rightarrow 0$.

\begin{figure}[H]
\begin{center}
\includegraphics[width=\linewidth, width=14cm]{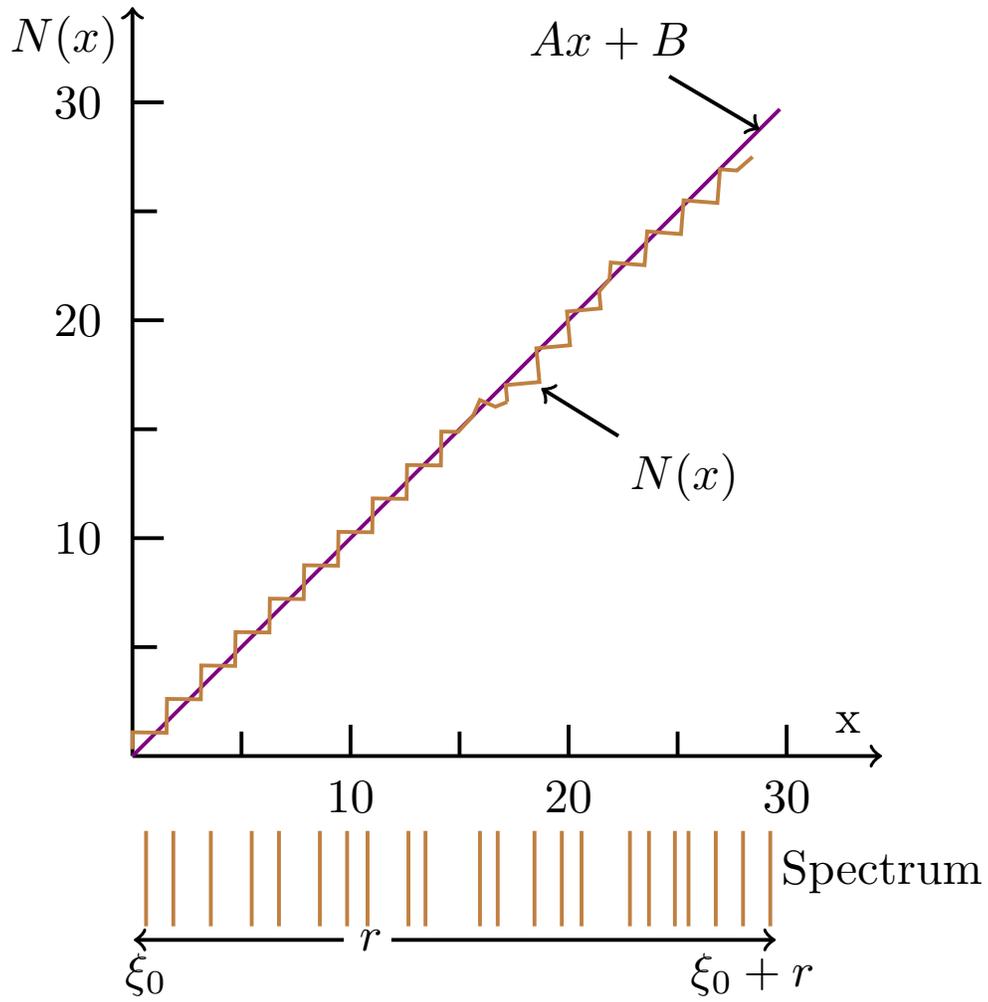}
\caption{Schematic diagram for calculation of $\Delta_{3}$.}\label{fig:DELTA3}
\end{center}
\end{figure}

Another frequently used fluctuation measure is the $\Delta_{3}$ statistic \cite{FJDY4}. For an interval $[\xi_{0}, \xi_{0}+r]$, we define the staircase function $N(x)$, which is the number of levels $\leq \xi_{0}+x$. The quantity $\Delta_{3}$ is the least square deviation of $N(x)$ from a best-fit straight line $Ax+B$ (see Fig.~\ref{fig:DELTA3}):
\begin{equation}\label{Eq.24}
\Delta_{3}(r)= \frac{1}{r}~_{\mbox{A,B}}^{\mbox{Min}}~\int_{\xi_{0}}^{\xi_{0} +r} \left[ N(x)-Ax-B\right]^{2} dx.
\end{equation}

Its average is related to $\Sigma^{2}$, and hence to $Y_{2}$, by
\begin{equation}\label{Eq.25}
\overline{\Delta}_{3}(r)= \frac{2}{r^{4}} \int_{0}^{r}\left(r^{3}-2r^{2}s+s^{3}\right) \Sigma^{2}(s)ds.
\end{equation}
Using $\Sigma^2 (s) = s$, we find $\overline{\Delta}_{3}(r)=r/15$ for the Poisson ensemble. Further, for the classical ensembles with $r \geq 1$,
\begin{equation}\label{Eq.26}
\overline{\Delta}_{3}(r)= \frac{1}{2}\Sigma^{2}(r)-\frac{9}{4\beta \pi^{2}}.
\end{equation}

To compute $\Delta_{3}$, we consider an ordered sequence of eigenvalues $x_{1},\cdots,x_{n}$ in the interval $\left[ \xi_{0}, \xi_{0}+r \right]$ containing $n$ eigenvalues. For this sequence, $\Delta_{3}$ is given by \cite{BG}
\begin{align}\label{Eq.27}
\Delta_{3}(\xi_{0}; r)= \frac{n^{2}}{16}- \frac{1}{r^{2}}\left(\sum_{j=1}^{n}\tilde{x}_{j} \right)^{2}+\frac{3n}{2r^{2}} \sum_{j=1}^{n}\tilde{x}_{j}^{2} \nonumber \\
~~~-\frac{3}{r^{4}} \left( \sum_{j=1}^{n}\tilde{x}_{j}^{2} \right)^{2} + \frac{1}{r}\left[ \sum_{j=1}^{n}(n-2j+1)\tilde{x}_{j} \right],
\end{align}
where $\tilde{x}_{j}= x_{j}-(\xi_{0}+\frac{r}{2})$. Note that the number of eigenvalues $n$ has the average value $r$. $\Delta_{3}$ is the ensemble average. Fluctuation measures derived from higher-order correlation functions have also been used in numerical analysis \cite{BHP, HPB}.

The above formulas apply to energy level spectra obtained from a single random matrix ensemble. We can also consider a direct product of $l$ independent random matrices with dimensions $f_{j}N$ (where $j=1, \cdots, l$ with $\sum_{j=1}^l  f_{j}=1$). This corresponds to the superposition of $l$ independent spectra. In this case, the fluctuation properties are generalized as
\begin{equation} \label{Eq.28}
Y_{2}^{(\text{mix})}(r)= \sum_{j=1}^{l}(f_{j})^{2}Y_{2}^{(j)}(f_{j}r),
\end{equation}
and the number variance is given by
\begin{equation}\label{Eq.29}
\Sigma^{2,(\text{mix})}(r)= \sum_{j=1}^{l} \Sigma^{2,(j)}(f_{j}r).
\end{equation}
Here, $Y_{2}^{(j)}$ and $\Sigma^{2,(j)}(r)$ are the two-level cluster function and number variance of the $j$-th component, respectively. There is a similar result for $\bar{\Delta}_{3}$, which can be derived from Eq.~(\ref{Eq.25}). In Eqs~(\ref{Eq.28})-(\ref{Eq.29}), for $l$ large and $f_{j} \sim \frac{1}{l}$, one can easily prove that $Y_{2}(r)\rightarrow 0$, and $\Sigma^{2}(r) \rightarrow r$, corresponding to Poisson spectra. The nearest-neighbor spacing distribution for the superposition of the $l$ independent sequences $p_{0}^{(j)}(s)$ is \cite{ML}
\begin{equation}\label{Eq.30}
p_{0}^{\text{(mix)}}(s)= \frac{d^{2}E_{0}^{\text{(mix)}}(s)}{ds^{2}}.
\end{equation}
Here,
\begin{align}\label{Eq.31}
E_{0}^{\text{(mix)}}(s)& =\prod_{j=1}^{l}E^{(j)}(s)(f_{j}s) , \nonumber \\
E^{(j)}(s)&= \int_{s}^{\infty}\left[ 1-F^{(j)}(r)\right] dr, \nonumber \\
F^{(j)}(s)&= \int_{0}^{s} p_{0}^{(j)}(r)dr .
\end{align}

For $l=2$, i.e., superposition of two independent sequences, with $f_1=f_2=1/2$ and $p_{0}^{(1)}(r)=p_{0}^{(2)}(r)=p_{0}(r)$ from Eq.~(\ref{Eq.20}) for $\beta = 1$, we obtain
\begin{eqnarray}\label{Eq.32}
p_{0}^{\text{(mix)}}(s)= \frac{1}{8} e^{- \pi s^2/8} \left[ s \pi \exp \left(\frac{\pi s^2}{16} \right) \text{erfc}\left(\frac{\sqrt{\pi}s}{4}\right) +4 \right] . \nonumber \\
\end{eqnarray}
The number variance for $l=2$, with  $f_1=f_2=1/2$ and $\Sigma^{2,(1)}(r)=\Sigma^{2,(2)}(r)= \Sigma_{\beta=1}^{2}(r)$ from Eq.~(\ref{Eq.18}), is obtained from Eq.~(\ref{Eq.29}) as
\begin{equation}\label{Eq.33}
\Sigma^{2,(\text{mix})}(r)= 2 \Sigma_{\beta=1}^{2}\left( \frac{r}{2}\right).
\end{equation}
Similarly, for the superposition of two GOE, the $Y_2^{(\text{mix})}$ function from Eq.~(\ref{Eq.28}) is
\begin{equation}\label{Eq.33c}
Y_2^{(\text{mix})}(r)= \frac{1}{2} Y_2^{\beta=1} \left( \frac{r}{2} \right).
\end{equation}

Before concluding this section, it is useful to describe numerical procedures for generating matrix ensembles (as in Eq.~(\ref{Eq.1})) and their eigenvalue spectra as in Eqs.~(\ref{Eq.2})-(\ref{Eq.4}). For generating matrices, the Gaussian cases are easiest as only Gaussian random numbers are involved (see the discussion after Eq.~(\ref{Eq.1})). The uniform circular ensembles for $\beta=2$ can be generated by constructing $N$ complex random vectors and orthogonalizing them. For $\beta=1, 4$ one can construct the matrices from the $\beta=2$ matrices $U$ by calculating $UU^{T}$ for $\beta=1$ and $UU^{D}$ for $\beta=4$.\par 

For more general potentials $V(A)$, one can use the Monte Carlo (MC) method to generate matrix ensembles as in Eq.~(\ref{Eq.1}) and spectra as in Eqs.~(\ref{Eq.2})-(\ref{Eq.4}). We confine our discussion to the case of eigenvalue ensembles.

In the \textit{linear case} \cite{GPPS}, we take a set of $N$ eigenvalues  $\left(x_{1},\cdots,x_{N}\right)$ ordered sequentially on a real line with fixed boundaries. The boundaries are chosen such that the probability of finding an eigenvalue outside the range is negligible. A stochastic move assigns, to any randomly chosen $x_{k}$, the new position $x^{\prime}_{k}$ between $(x_{k-1},x_{k+1})$ with a uniform probability. The move is accepted with a probability $\exp(-\beta \triangle W)$, where $\Delta W$ is the change in the potential after the stochastic move. Time is measured in units of Monte Carlo steps (MCS), with 1 MCS corresponding to $N$ attempted eigenvalue moves.

In the \textit{circular case} \cite{SANDP}, we take a set of $N$ eigenvalues $\left(e^{i\theta_{1}},\cdots,e^{i\theta_{N}}\right)$ ordered sequentially on the unit circle. Again, a stochastic move considers a randomly chosen eigenangle $\theta_{j}$, and assigns it the new position $\theta^{\prime}_{j}$ between $(\theta_{j-1},\theta_{j+1})$ with a uniform probability. After each eigenangle  movement we apply periodic boundary conditions, i.e., $\theta^{\prime}_{j}$ is computed modulo $2\pi$. This clearly respects the original order of their positions. Again the move is accepted with a probability $\exp(-\beta \triangle W)$, where $\Delta W$ is the change in the potential after the stochastic move. After 1 MCS, the new state with eigenangle positions $\left(\theta^{\prime}_{1},\cdots,\theta^{\prime}_{N}\right)$ always has $\theta^{\prime}_{i}< \theta^{\prime}_{j}$ for $i<j$.

\section{Statistical Properties of Eigenvectors}\label{S4}

The eigenvectors of random matrices are useful in the study of fluctuations of transition widths, and expectation values. The eigenvectors are random subject to the conditions of orthonormality. We focus on a typical eigenvector $u$ which consists of $N$ elements $u_{j}$. Each element has $\beta$ components $\lbrace u_{j}^{(\gamma)}: \gamma=0,\cdots,\beta-1\rbrace$ \cite{BRODY}. The jpd of the components and elements of a single eigenvector $\{u_{j}^{(\gamma)}\}$ is given by
\begin{equation}\label{Eq.34}
Q\left( \{ u_{j}^{(\gamma)} \} \right) = \pi^{-\frac{\beta N}{2}}~\Gamma\left(\frac{\beta N}{2}\right)\delta\left(\sum_{j=1}^{N}|u_{j}|^{2}-1\right).
\end{equation}
Here, the expression $|u_{j}|^{2}$ refers to the sum of absolute squares of $u_{j}^{(\gamma)}$.

\begin{figure}[H]
\begin{center}
\includegraphics[width=\linewidth, width=16cm]{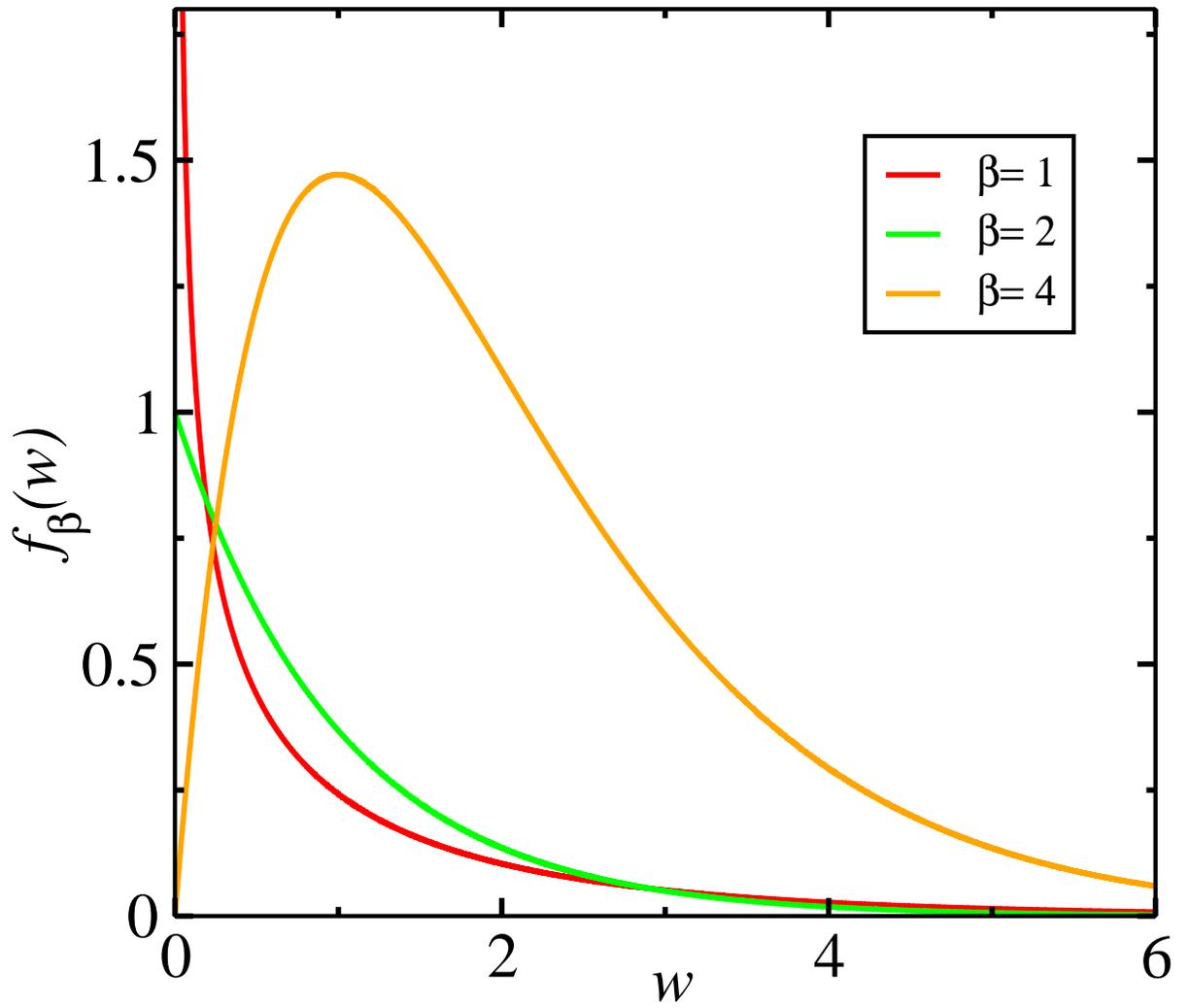}
\caption{Plot of $f_{\beta}(w)$ vs. $w$ for $\beta=1, 2 ,4$.}\label{fig:EIGENVECTOR}
\end{center}
\end{figure}

The distribution of a single element of an eigenvector is given by
\begin{equation}\label{Eq.35}
Q(u_{j})= \frac{\pi^{-\frac{\beta}{2}}\Gamma\left( \frac{\beta N}{2}\right)} {\Gamma\left( \frac{\beta (N-1)}{2}\right) }\left(1-|u_{j}|^{2}\right)^{\frac{\beta (N-1)}{2}-1}.
\end{equation}
For large $N$, the $u_{j}^{(\gamma)}$ become independent Gaussian variables with mean $0$ and variance $1/(\beta N)$. For a given $j$, the variable $w=N|u_{j}|^{2}$ has a $\chi_{\beta}^{2}$-distribution with mean $1$, variance $2/\beta$:
\begin{equation}\label{Eq.36}
f_{\beta}(w)=\frac{(\frac{\beta}{2})^{\frac{\beta}{2}}}{\Gamma(\frac{\beta}{2})}  w^{\frac{\beta}{2}-1} \exp\left( -\frac{\beta w}{2}\right).
\end{equation}
This distribution function is plotted in Fig.~\ref{fig:EIGENVECTOR} for $\beta = 1,2,4$. For $\beta=1$, it is known as the Porter-Thomas distribution in nuclear physics:
\begin{equation}\label{Eq.37}
f_{1}(w) = \frac{1}{\sqrt{2\pi}}~w^{-\frac{1}{2}} \exp(-w/2).
\end{equation}
This distribution is realized in physical systems as follows. The transition matrix elements connecting the states in a narrow energy band to a lower energy state may be treated as independent zero-centered Gaussian random variables. Then, Eq.~(\ref{Eq.34}) describes the distribution of the transition widths ($\propto$ absolute square of the matrix elements) normalized to unity.

\section{Application to Nuclear Spectra}\label{S5}

As mentioned earlier, Wigner introduced RMT to study slow neutron resonances of heavy nuclei. Levels with fixed angular momentum and parity were used (typically, \ce{\frac{1}{2}^{+}} levels). The subsequent mathematical developments motivated numerous experiments on nuclear spectra, mostly performed at Columbia university. There were also some experiments on proton resonances in light nuclei. The utility of GOE in understanding nuclear spectra was broadly established by these comparisons. In the early 1980's, the same data was reanalyzed with some new features \cite{HPB, BHPNDE, BHP}: \\
(a) The nuclear spectra from different experiments were combined to enhance the statistics. The combined data was referred to as ``nuclear data ensemble'' (NDE). \\
(b) More sophisticated statistical measures were used to analyze the spectral fluctuations. \\
(c) The sample errors were carefully calculated and used to quantify the level of agreement.

\begin{figure}[H]
\begin{center}
\includegraphics[width=\linewidth, width=10cm]{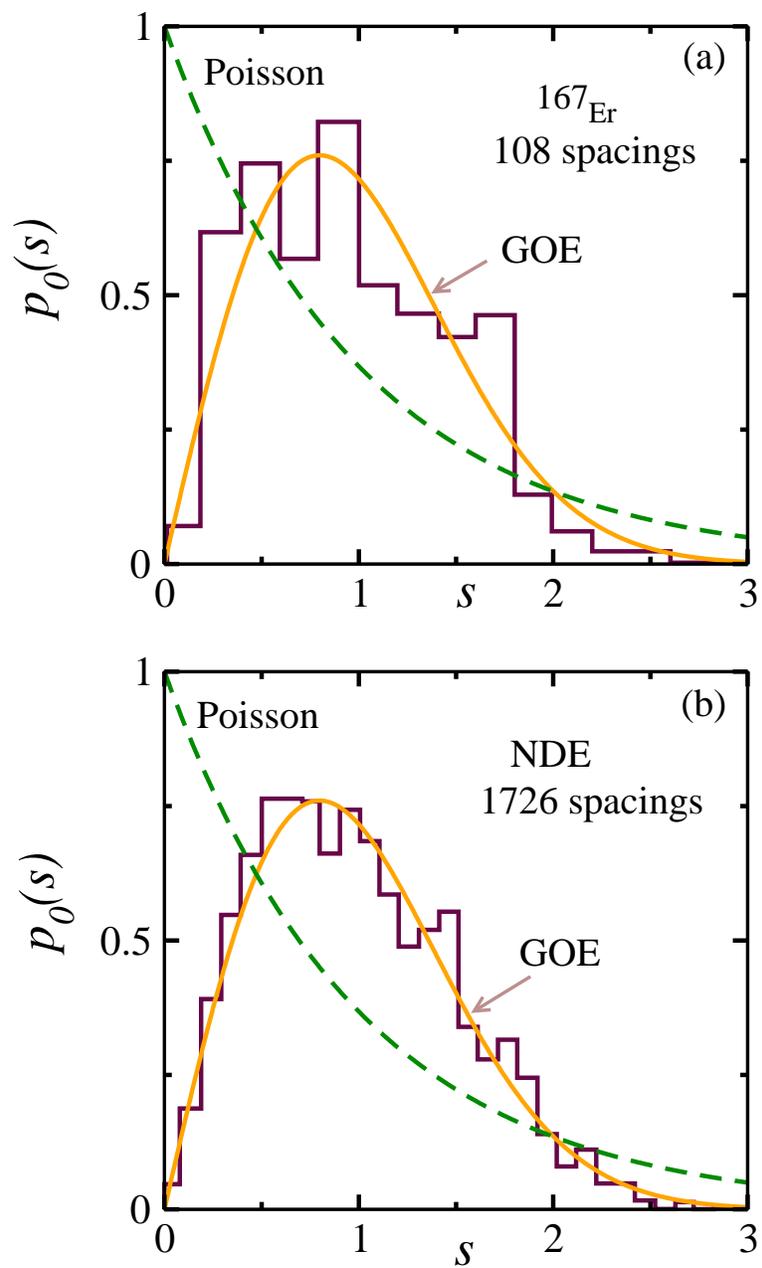}
\caption{Nearest-neighbor spacing distribution for (a) \ce{^{167}Er}, (b) Nuclear data ensemble (NDE)} \label{fig:NDE1}
\end{center}
\end{figure}

\begin{figure}[H]
\begin{center}
\includegraphics[width=\linewidth, width=14cm]{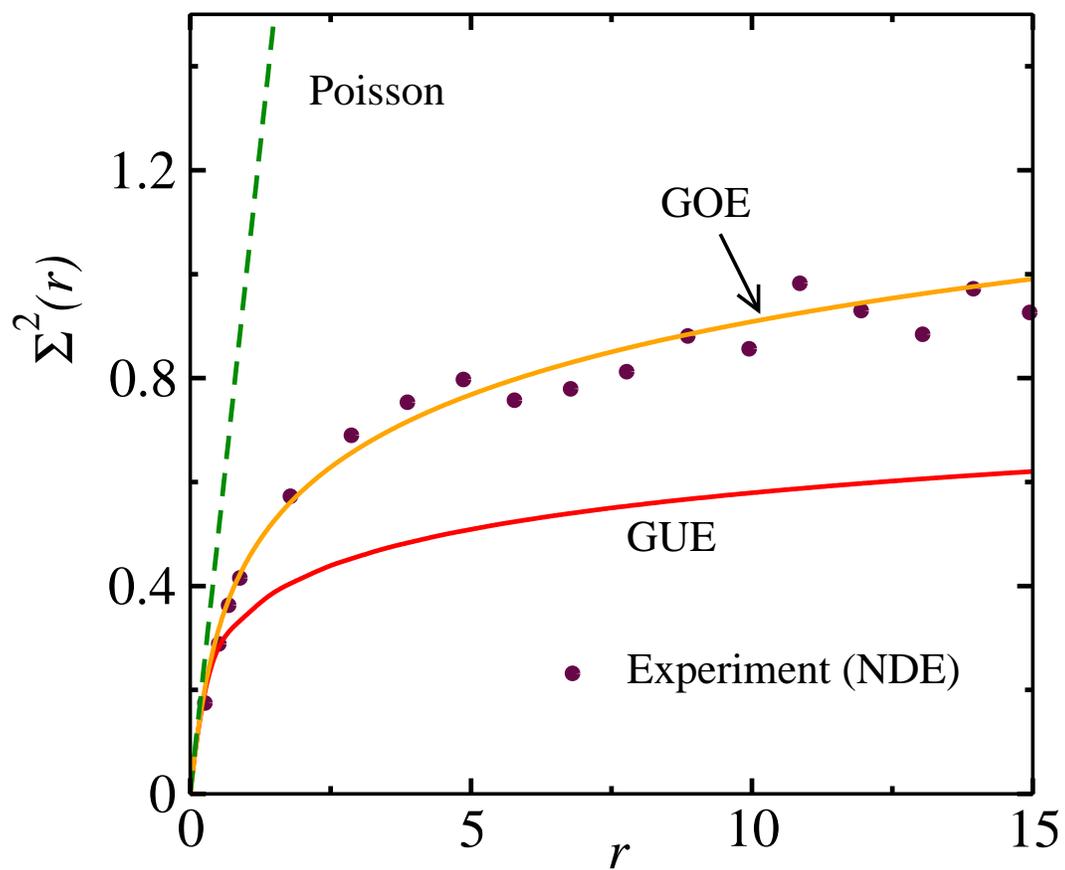}
\caption{Number variance $\Sigma^{2}(r)$  vs. $r$ for the NDE.} \label{fig:NDE2}
\end{center}
\end{figure}

\begin{figure}[H]
\begin{center}
\includegraphics[width=\linewidth, width=14cm]{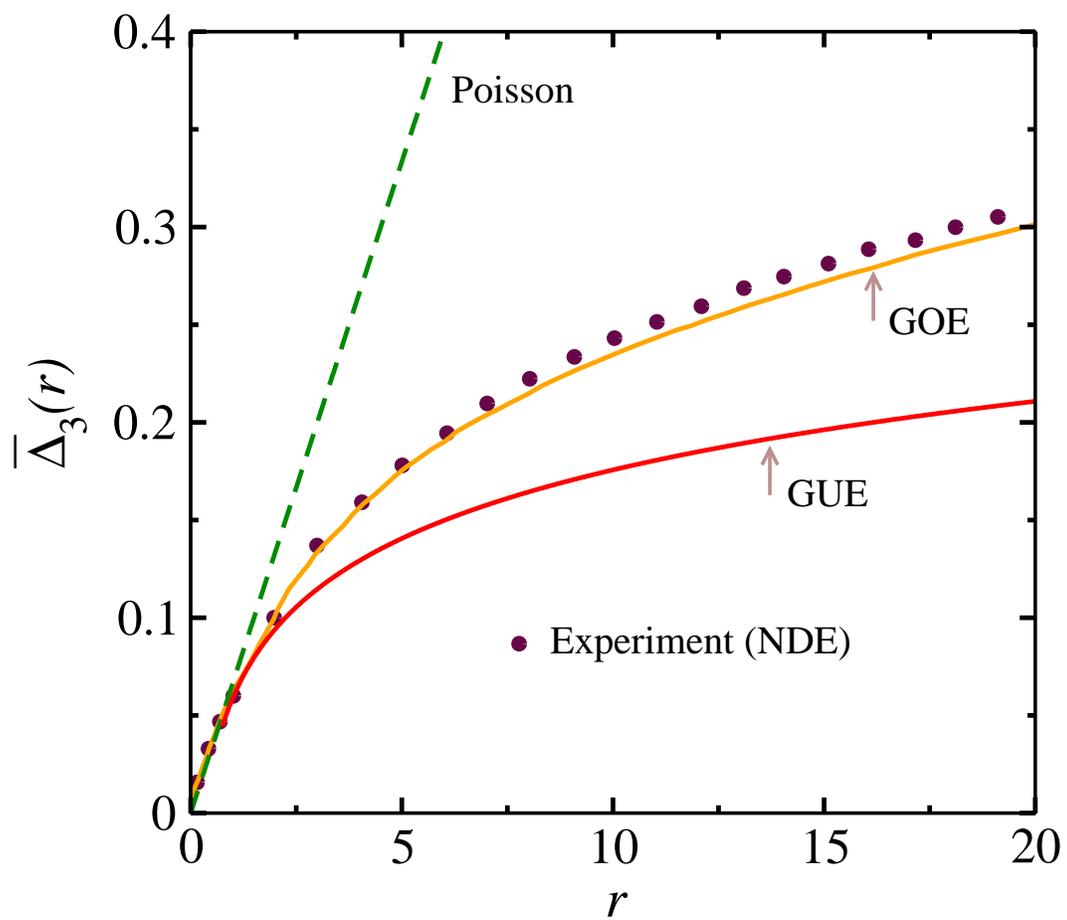}
\caption{$\bar{\Delta}_{3} (r)$ vs. $r$ for the NDE.} \label{fig:NDE3}
\end{center}
\end{figure}

\begin{figure}[H]
\begin{center}
\includegraphics[width=\linewidth, width=10cm]{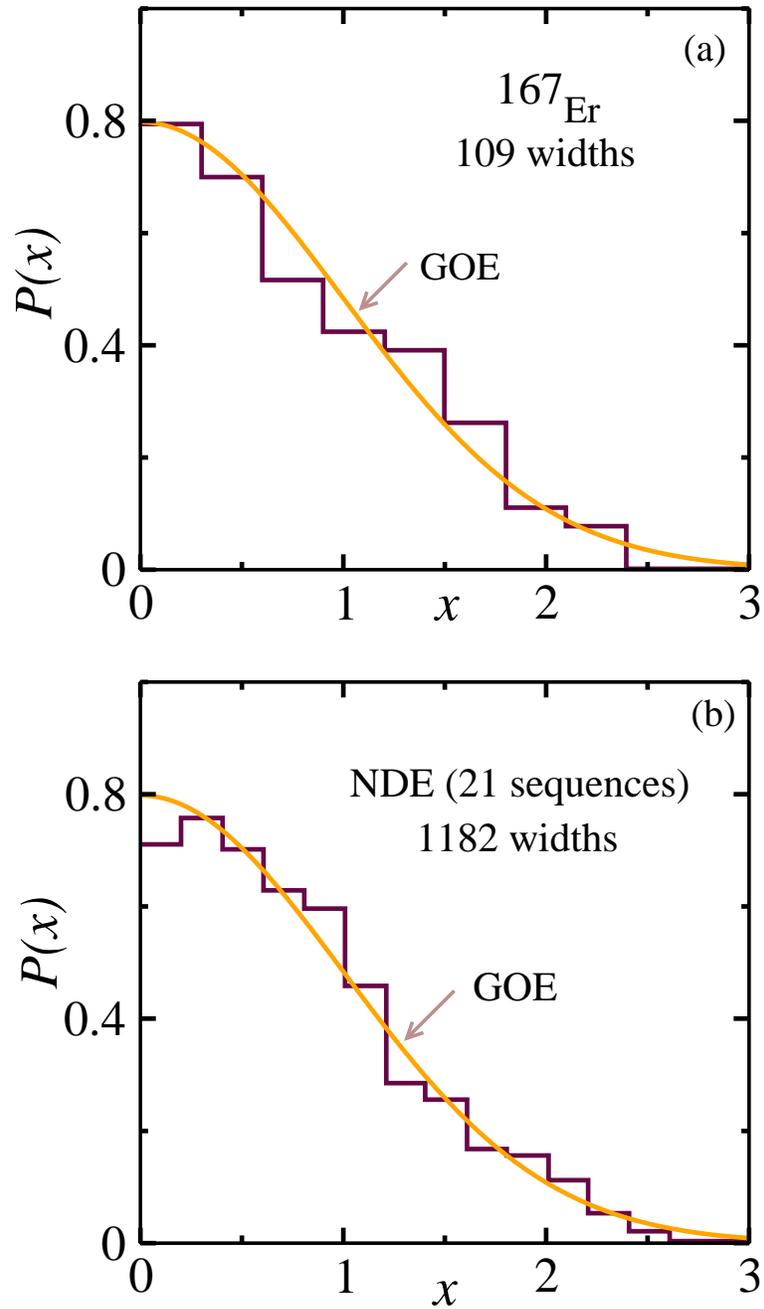}
\caption{Distribution of the square roots of transition widths for (a) \ce{^{167}Er}, and (b) the NDE. The solid line denotes the Gaussian density: $P(x)=\sqrt{(2/\pi)}\exp(-x^{2}/2)$.} \label{fig:NDE4}
\end{center}
\end{figure}

In this section, we will give a brief review of the NDE analysis. In this analysis, the NDE consisted of $1762$ resonance energies corresponding to $36$ sequences of $32$ different nuclei. Note that, for a single nucleus, the fluctuation measures are calculated as spectral averages. For the NDE, there is further averaging over the ensemble of nuclei. We show the nearest-neighbor spacing histogram of \ce{^{167}Er} (obtained from n $+$ \ce{^{166}Er}) in Fig.~\ref{fig:NDE1}~(a), and of the NDE in Fig.~\ref{fig:NDE1}~(b). One can see that the quality of agreement with GOE improves considerably when the spectra of many nuclei are combined. In Figs.~\ref{fig:NDE2} and \ref{fig:NDE3}, we show $\Sigma^{2}(r)$ and $\overline{\Delta}_{3}(r)$ for the NDE. Again, very good agreement is found with GOE. In Fig.~\ref{fig:NDE4}, we show histograms for the distribution of square roots of transition widths for (a) \ce{^{167}Er}, and (b) the NDE consisting of $1182$ widths of $21$ sequences. The result is compared with the Gaussian distribution. (Note that the Gaussian distribution arises from the Porter-Thomas distribution in Eq.~(\ref{Eq.37}) for the variable $x = \sqrt{w}$.)

Finally, we discuss the correlation between the energy levels $\{ x_j \}$ and the corresponding normalized widths $\{ w_j \}$. RMT predicts the width and energy level fluctuations to be independent. We define the correlation coefficient $r$ (for each sequence) as
\begin{equation}\label{Eq.38}
r= \frac{1}{N}\sum_{j=1}^{N} \left( \frac{w_{j}-\langle w \rangle}{\sigma_{w}}\right) \left(  \frac{x_{j}-\langle x_{j}\rangle}{\sigma_{x}} \right) ,
\end{equation}
where $N$ is the number of levels in the sequence. Here $\langle w \rangle$ and $\sigma_{w}^{2}$ are the mean and variance of $\{ w_j \}$, and
\begin{equation}\label{Eq.39}
\langle x_{j}\rangle = Aj+ B, ~~~ \sigma_{x}^{2}=\frac{1}{N}\sum_{j=1}^{N}\left( x_{j}-\langle x_{j} \rangle \right)^{2}.
\end{equation}
The parameters $A$ and $B$ are calculated by minimizing the expression for $\sigma_{x}^{2}$. For the NDE, one evaluates the average of the correlation coefficient $r$, weighted according to the size of the sequence. For $1182$ widths with $21$ sequences, one obtains $r(NDE)=0.017$, confirming the independence of widths and energy level spectra.

\section{Quantum Chaos and Random Matrices}\label{S6}

It is relevant to ask why the GOE works so well for nuclear spectra. We emphasize that the system's physical symmetries are important in determining the relevant level statistics. Thus, for example, the GUE is not encountered in TRI systems, e.g., nuclear spectra. Moreover the quantum numbers should not be mixed as this leads to superposition of independent spectra leading to Poisson statistics (see Sec.~\ref{S3}). Apart from nuclear spectra, random matrix statistics is also found in complex atomic and molecular spectra. Rosenzweig and Porter \cite{RP} have shown that levels having the same quantum numbers show level repulsion, and the spacing distribution follows Wigner's prediction. More generally, it has emerged that quantum chaotic systems, viz., quantum systems whose classical analogs are chaotic, follow random matrix statistics. In this section, we will provide a brief overview of some important developments in quantum chaos.

In early work, Percival \cite{PERCIVAL} introduced the terminology `regular and irregular' spectra, which arise in quantum analogs of integrable and chaotic classical systems, respectively. In 1977, Berry and Tabor showed that the regular spectra follow Poisson statistics and display level clustering \cite{BT}. Their proof used semiclassical quantization  of integrable systems.

Subsequently, Mcdonald-Kaufman \cite{MK} and Berry \cite{BERRY1} considered stadium and Sinai billiards respectively, and showed that their quantum spectra display level repulsion. Classically, a billiard system refers to a free particle moving in a two-dimensional region, obeying classical reflection rules at the boundary of the region. For the corresponding quantum billiard, the eigenfunctions are solutions of the free-particle Schrodinger equation which vanish at the boundary of the region. The energy levels are discrete.

\begin{figure}[H]
\begin{center}
\includegraphics[width=\linewidth, width=8cm]{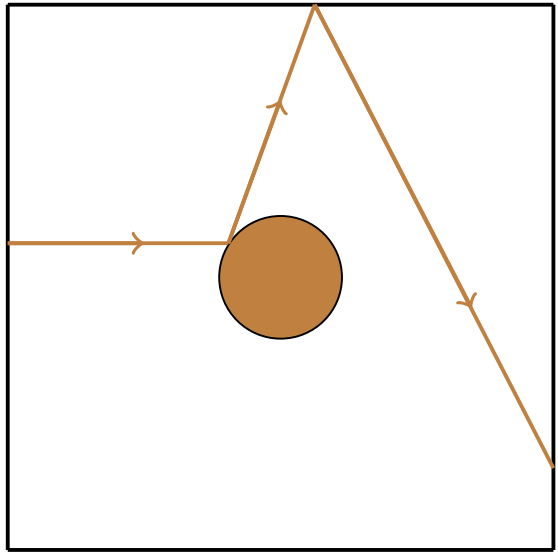}
\caption{Sinai billiard.}\label{fig:SINAI}
\end{center}
\end{figure}

\begin{figure}[H]
\begin{center}
\includegraphics[width=\linewidth, width=8cm]{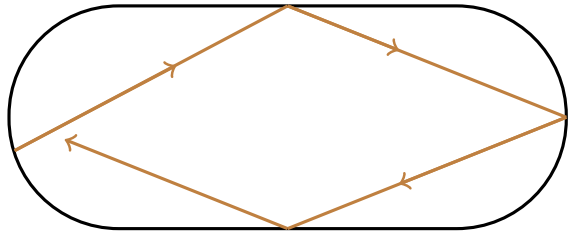}
 \caption{Stadium billiard.}\label{fig:STADIUM}
\end{center}
\end{figure}

The Sinai billiard is a square region with a circular disk at the center (Fig.~\ref{fig:SINAI}). The stadium billiard is a region bounded by two half-circles joined by a rectangle (Fig.~\ref{fig:STADIUM}). Both these systems exhibit strong classical chaos. Quantum billiards are experimentally realized as quantum dots in both mesoscopic and nanoscopic systems. For quantum studies, desymmetrized billiards were used in both cases so that the spectra corresponded to the same quantum numbers.

In 1984, Bohigas et al. \cite{BGS} considered the same quantum billiards again. However, they discarded the low-lying energy levels to obtain much sharper correspondence with GOE. The corresponding spectra accurately displayed the properties of level repulsion and spectral rigidity found in GOE. The deep connection between quantum chaos and RMT is now known as the BGS conjecture. Seligman et al. \cite{SVZ} showed GOE statistics in a system of coupled quantum nonlinear oscillators. Berry and Robnik \cite{BR} found GUE statistics in the Aharonov-Bohm billiard system, where TRI is broken by the introduction of a magnetic field. Another important study in this context is the spectral analysis of the anisotropic Kepler problem \cite{WM}. Berry gave a semiclassical theory of spectral rigidity, using Gutzwiller's periodic orbit theory \cite{GUTZ}. 

Finally, we mention the quantum version of periodically-kicked rotors, also referred to as quantum kicked rotors (QKR). In this system, both GOE and GUE statistics are found \cite{FM1}. In Secs.~\ref{S8} and \ref{S9}, we make a detailed study of the classical and quantum properties of kicked rotor systems. We remark that the QKR has received much attention in the literature because its numerical implementation is much easier than that of the quantum billiard.

\section{Quantum integrable systems and Poisson statistics}\label{S7}

\begin{figure}[H]
\begin{center}
\includegraphics[width=\linewidth, width=10cm]{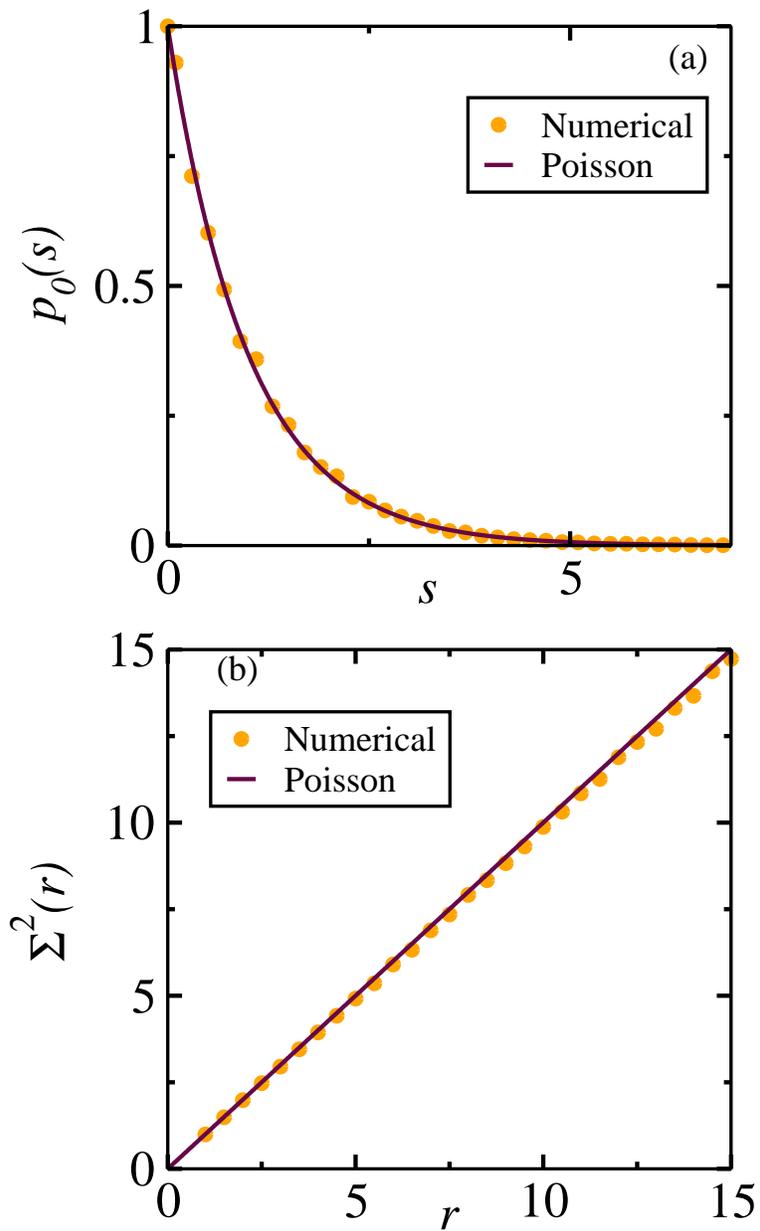}
\caption{Spectral analysis of the system corresponding to Eq.~(\ref{Eq.40}). (a) Nearest-neighbor spacing distribution $p_0(s)$, and (b) Number variance $\Sigma^2 (r)$.}\label{fig:INTEGRABLE}
\end{center}
\end{figure}

As mentioned above, Poisson statistics is encountered in quantum counterparts of integrable systems. We give here two simple examples to illustrate this. First we consider the case where the eigenvalue spectrum can be written as 
\begin{equation}\label{Eq.40}
x_{j}=j^{2} \alpha, \quad (\mbox{mod 1}),
\end{equation}
where $\alpha$ is an irrational number. In this system, the average density $\bar{\rho}(x)$ is constant. In Fig.~\ref{fig:INTEGRABLE}, we show that the spacing distribution [$p_{0}(s)$] and the number variance [$\Sigma^{2}(r)$] obey Poisson statistics. This spectrum can arise in a two-dimensional system with harmonic binding in one direction and a rigid wall in the other direction.

The system of harmonic oscillators is a major exception to the ``integrability implies Poisson'' rule. For example, consider the spectrum 
\begin{equation}
x_{j}=j\alpha, \quad (\mbox{mod 1}),
\end{equation}
which arises in a system of two-dimensional harmonic oscillators. This system does not exhibit Poisson or any other universal statistics \cite{PR, PBG}.

Our second example of Poisson statistics is the rectangular billiard \cite{CCG}. The eigenvalues can be written as
\begin{equation}\label{Eq.41}
E_{n_{1}, n_{2}}= \alpha_{1}n_{1}^{2}+\alpha_{2}n_{2}^{2},
\end{equation}
where $n_{1}, n_{2}= 1, 2, 3, \cdots$ and $\alpha_{1}, \alpha_{2}$ contain information about the lengths of the sides. In this case also, the level density is constant. When $\alpha_{1}, \alpha_{2}$ are not rationally connected, Poisson statistics is obtained.

\section{Classical kicked rotor}\label{S8}

Periodically kicked rotors have been widely used in studies of classical and quantum chaos \cite{HS, CC, FM2, PF,CMI}. In this section, we briefly review the classical kicked rotor.

The Hamiltonian of a periodically kicked rotor can be written as 
\begin{equation}\label{Eq.42}
H = \frac{p^{2}}{2} + V(\phi)\sum_{n=-\infty}^{\infty}\delta\left(t-n \right).
\end{equation}
Here, $\phi$ is the angle of rotation and $p$ is the angular momentum. Without loss of generality, we have chosen both the moment of inertia and the time period of kicking to be unity. The classical motion is such that there is free rotation in the time interval $n+1> t > n$. At each integer time $n$, the rotor is subjected to an impulse of magnitude $|V^{\prime}(\phi_{n})|$, where $\phi_{n}$ is the value of $\phi$ at $t=n$. Let us also define $p_{n}$ as the value of the momentum at $t=n-0$. Note that, at $t=n+0$, the momentum will be $p_{n+1}$. Thus, one can write the kicked rotor map as 
\begin{align}
&\phi_{n+1}= \phi_{n} + p_{n+1},\label{Eq.43}\\
&p_{n+1}= p_{n}-V^{\prime}(\phi_{n}).\label{Eq.44}
\end{align}

Consider the potential $V(\phi)= \alpha \cos \phi$, where $\alpha$ is the kicking parameter (also called ``the chaos parameter''). This gives the Chirikov standard map:
\begin{align}
&\phi_{n+1}=\phi_{n}+p_{n+1},    &      (\text{mod}~2\pi), \label{Eq.45}\\
&p_{n+1}= p_{n}+\alpha \sin \phi_{n}, &  (\text{mod}~2\pi). \label{Eq.46}
\end{align}
We have used the operation $(\text{mod}~2\pi)$ for $p$ also since the potential $V(\phi)$ is periodic.

\begin{figure}[H]
\begin{center}
\includegraphics[width=\linewidth, width=12cm]{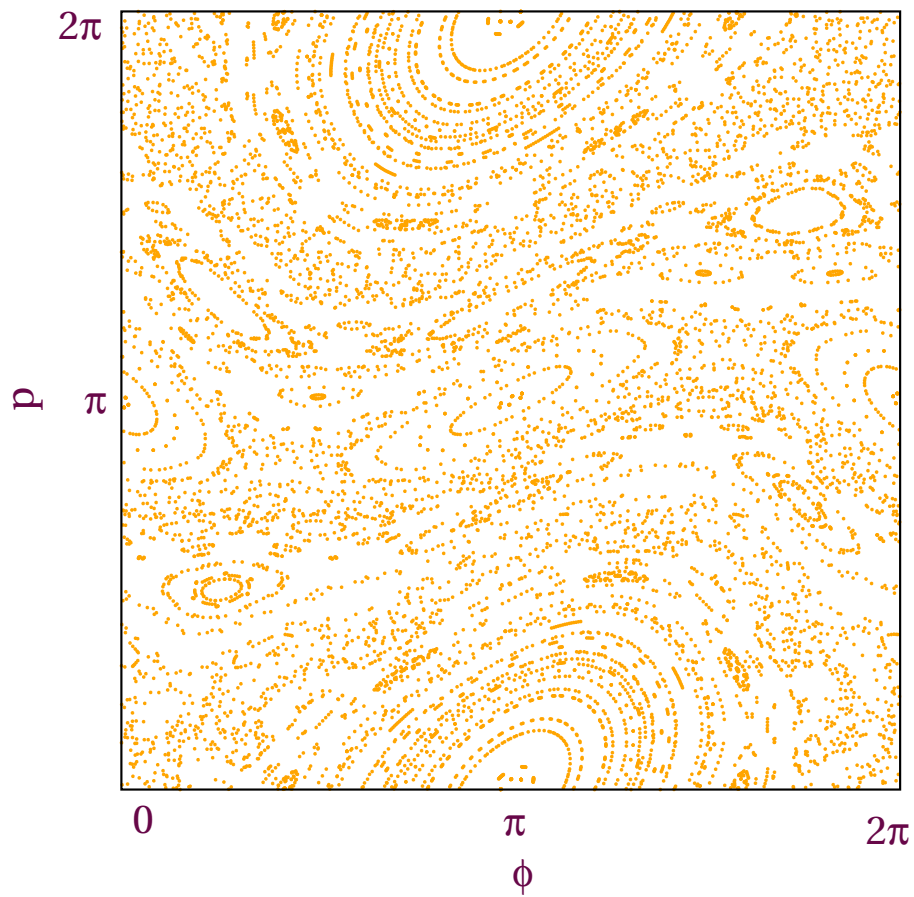}
\caption{Phase plot of the Chirikov map for (a) $\alpha=0$, (b) $\alpha=0.5$, (c) $\alpha=1$, and (d) $\alpha=10$. In each frame, we have taken 100 initial conditions with 250 iterations.} \label{fig:QKR-MAP}
\end{center}
\end{figure}

In Fig.~\ref{fig:QKR-MAP}, we show the phase plots of the Chirikov map in the $(p,\phi)$-plane for several values of $\alpha$.
The smooth curves indicate regular motion, whereas the dotted regions correspond to chaotic motion. The dashed regular curves in all figures correspond to periodic orbits. The different curves for regular motion arise from different initial conditions. On the other hand, even one initial condition can give rise to area-filling trajectories in the chaotic regime. For $\alpha=0$, we have only regular motion, whereas the trajectories are almost entirely chaotic for large $\alpha$. 

\section{Quantum kicked rotor}\label{S9}

We define QKR by using the unit time evolution operator $U$ to describe the evolution of the wave function. Because of the time-periodicity in the Hamiltonian, $U$ is time independent and is defined by
\begin{equation}\label{Eq.47}
\ket{\psi_{n+1}}= U\ket{\psi_{n}}.
\end{equation}
We consider $\ket{\psi_{n}}$ to be the state of the system at time $t=n-0$, as in the classical case. Eigenvalues of $U$ are of the form $e^{i\theta}$, where $\theta$ is called the eigenangle. 

In general, $U$ is an infinite-dimensional matrix. However, for the QKR with the cosine potential, one can construct finite-dimensional $U$ using the periodic boundary conditions of $p$ and $\theta$. We introduce parameters $\gamma$ and $\phi_{0}$ to describe time-reversal and parity breaking in the system by changing $p \rightarrow p+\gamma$ and $\phi \rightarrow \phi+\phi_{0}$ in Eq.~(\ref{Eq.42}). In the classical case, $\gamma$ and $\phi_{0}$ can be removed by a canonical transformation. However, in the quantum case, both parameters are very important and appear explicitly in the operators.

One can write $U=BG$, where $B$ is the operator for the evolution from time $t=n-0$ to $t=n+0$ and $G$ is the free evolution operator from $t=n+0$ to $t=n+1-0$. It can be shown that $B(\alpha)=\exp[-i\alpha \cos(\phi+\phi_{0})]$, and $G=\exp[-i(p+\gamma)^2/2]$, where we have set $\hbar=1$. For $N$-dimensional $U$, the matrix elements of $B$ in the position representation can be written as 
\begin{align}\label{Eq.48}
B_{jk}= \exp\left[ -i\alpha \cos\left( \frac{2\pi j}{N}+\phi_{0}\right)\right]\delta_{jk}.
\end{align}
Here, $j, k= -N^{\prime}, -N^{\prime}+1,\cdots, N^{\prime}$ with $N^{\prime}=(N-1)/2$. On the other hand, $G$ in the momentum representation can be written as
\begin{equation}\label{Eq.49}
G_{mn}=\frac{1}{N}\sum_{l=-N^{\prime}}^{N^{\prime}}\exp\left[ -i\left( \frac{1}{2}l^{2}-\gamma l-\frac{2\pi(m-n) l}{N}\right)\right],
\end{equation}
where $m, n= -N^{\prime},-N^{\prime}+1,\cdots, N^{\prime}$. Using the transformation properties between position and momentum representations, one can write the matrix elements of $U$ as
\begin{eqnarray}\label{Eq.50}
U_{jk}&=&\frac{1}{N} \exp \left[ -i\alpha \cos\left(\frac{2\pi j}{N} +\phi_{0}\right) \right]\times \nonumber \\ 
 &&\sum_{l=-N^{\prime}}^{N^{\prime}}\exp\left[-i\left( \frac{l^{2}}{2}-\gamma l -\frac{2\pi(j-k)l}{N} \right)  \right].
\end{eqnarray}

\section{Eigenvalue and Eigenvector fluctuations in QKR}\label{S10}

In this section, we will show representative results for eigenvalue and eigenvector statistics in the QKR using the fluctuation measures introduced in Sections \ref{S3}-\ref{S4}. We consider $\phi_{0} = \pi/(2N)$, corresponding to the case when parity is fully broken. (At some places, we will also consider the case $\phi_0 = 0$, corresponding to parity being preserved.) For $\gamma$, we choose two values: $\gamma=0.0$ and $\gamma=0.7$, corresponding to TRI and broken TRI respectively. These two values of $\gamma$ will give rise to COE (or GOE) and CUE (or GUE) statistics respectively. We consider $N=1000$. Since the spectra in the highly chaotic case ($\alpha$ very large) are known to become independent \cite{VKP} very rapidly with increasing $\alpha$, we generate spectra for $\alpha$ ranging from $10^{4}$ to $10^{6}$, in steps of $1000$. This gives us $1000$ independent spectra. As the density is uniform, the unfolded spectrum is obtained by multiplying the original eigenangles by $N/2\pi$.

\begin{figure}[H]
\begin{center}
\includegraphics[width=\linewidth, width=14cm]{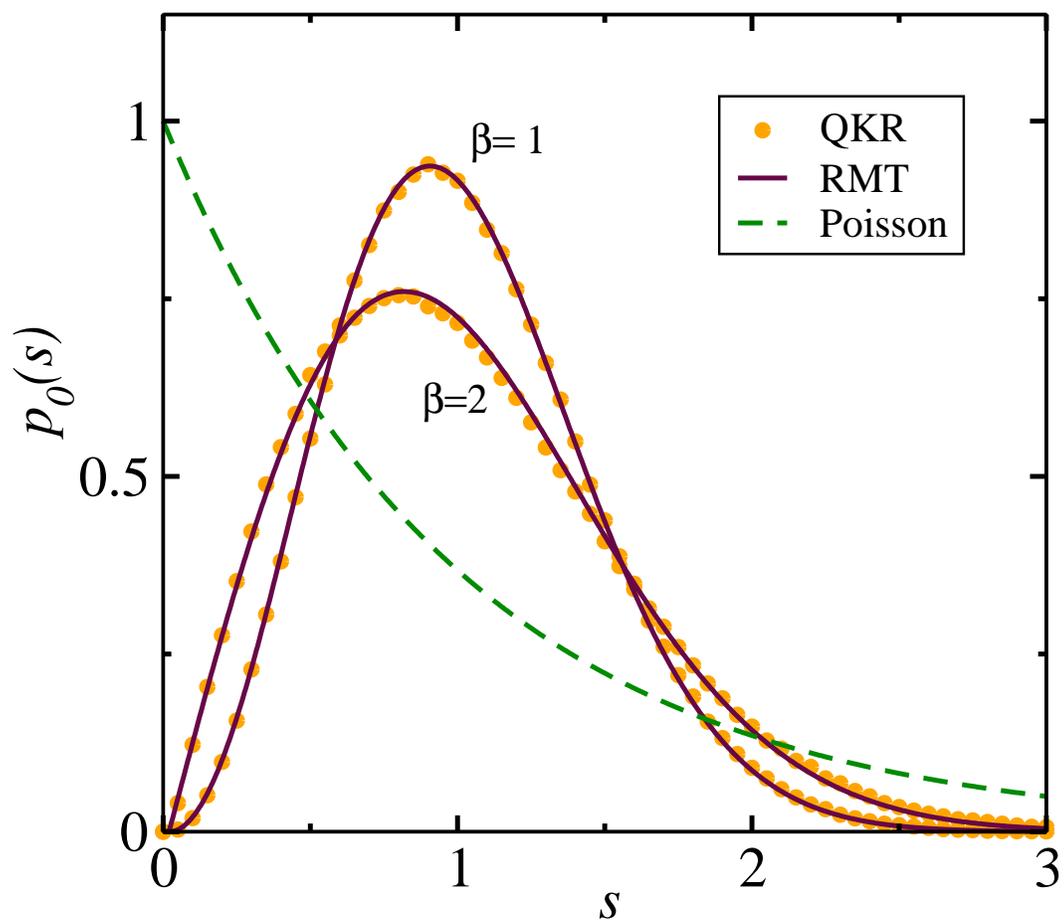}
\caption{Spacing distribution of QKR for $\gamma=0.0~(\beta=1)$ and $\gamma=0.7~(\beta=2)$.}\label{fig:QKR-NNSD}
\end{center}
\end{figure}

\begin{figure}[H]
\begin{center}
\includegraphics[width=\linewidth, width=14cm]{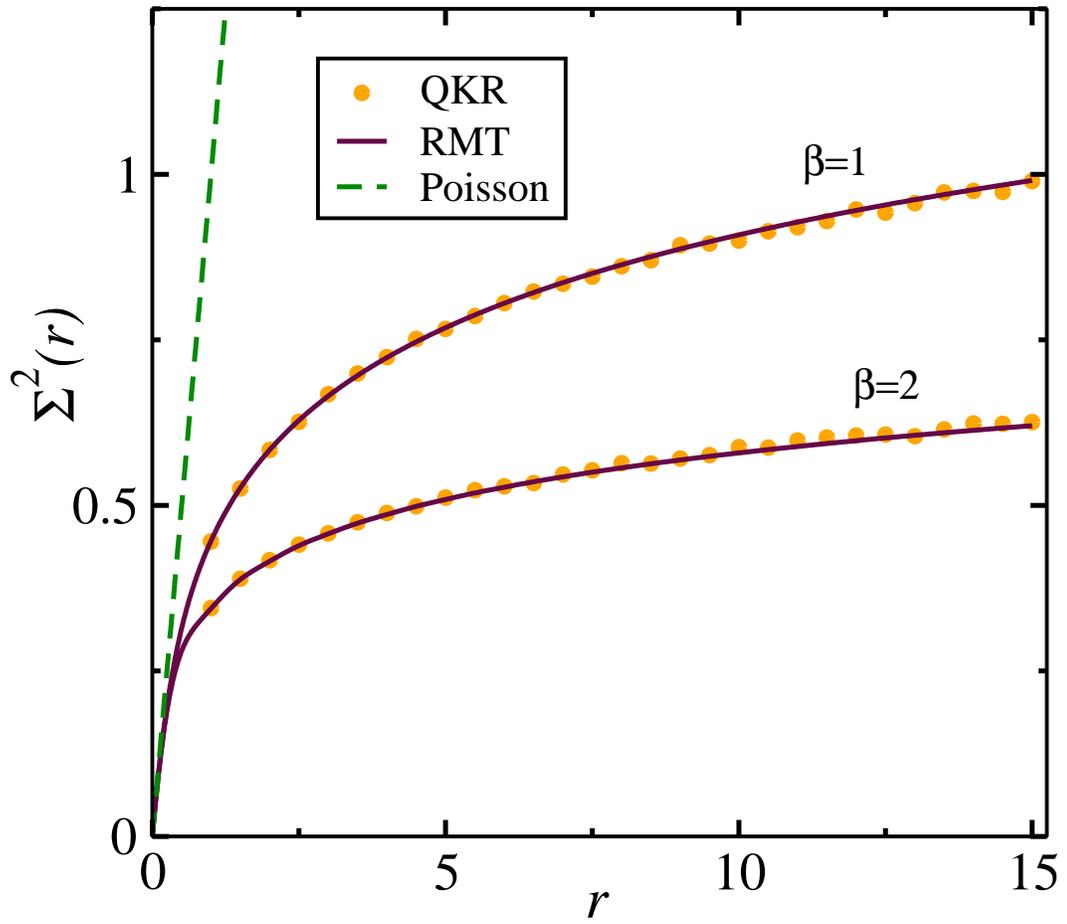}
\caption{Analogous to Fig.~\ref{fig:QKR-NNSD}, but for number variance.}\label{fig:QKR-NV}
\end{center}
\end{figure}

\begin{figure}[H]
\begin{center}
\includegraphics[width=\linewidth, width=10cm]{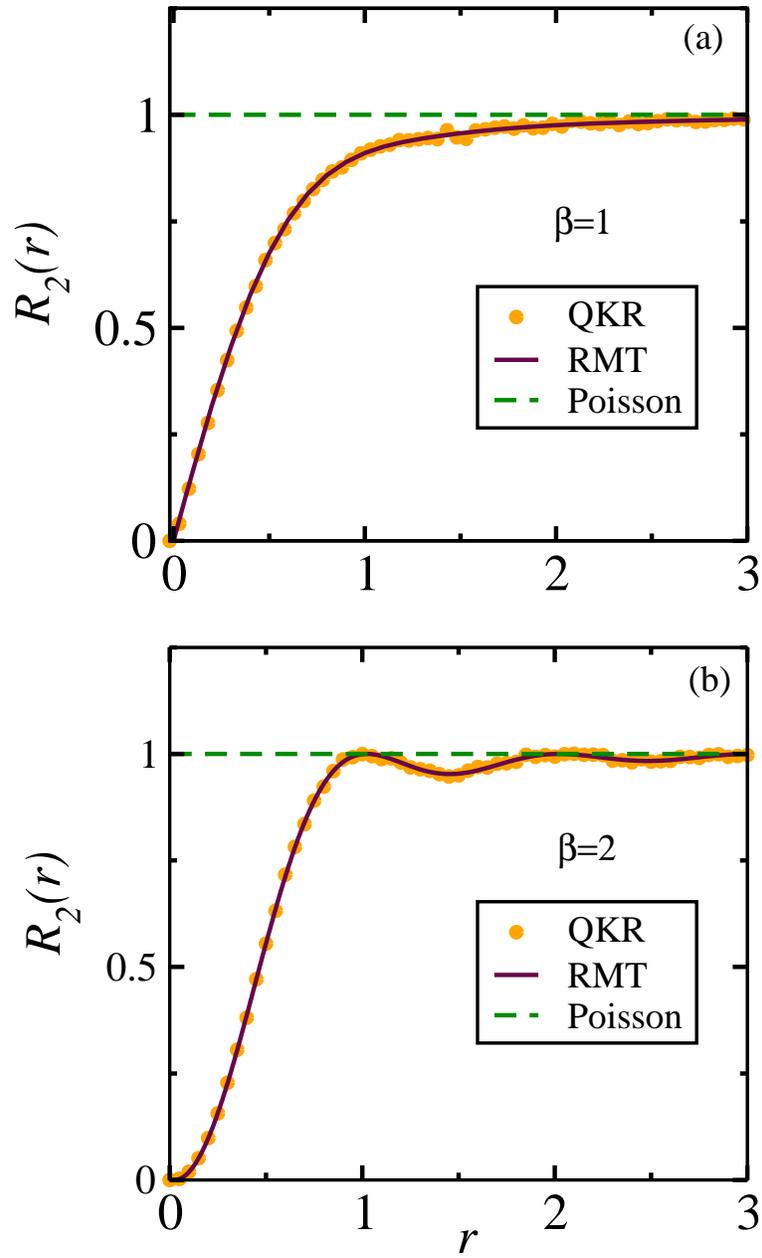}
\caption{Two-point correlation function of QKR for (a) $\gamma=0.0~(\beta=1)$, and (b) $\gamma=0.7~(\beta=2$).}\label{fig:QKR-R2}
\end{center}
\end{figure}

\begin{figure}[H]
\begin{center}
\includegraphics[width=\linewidth, width=10cm]{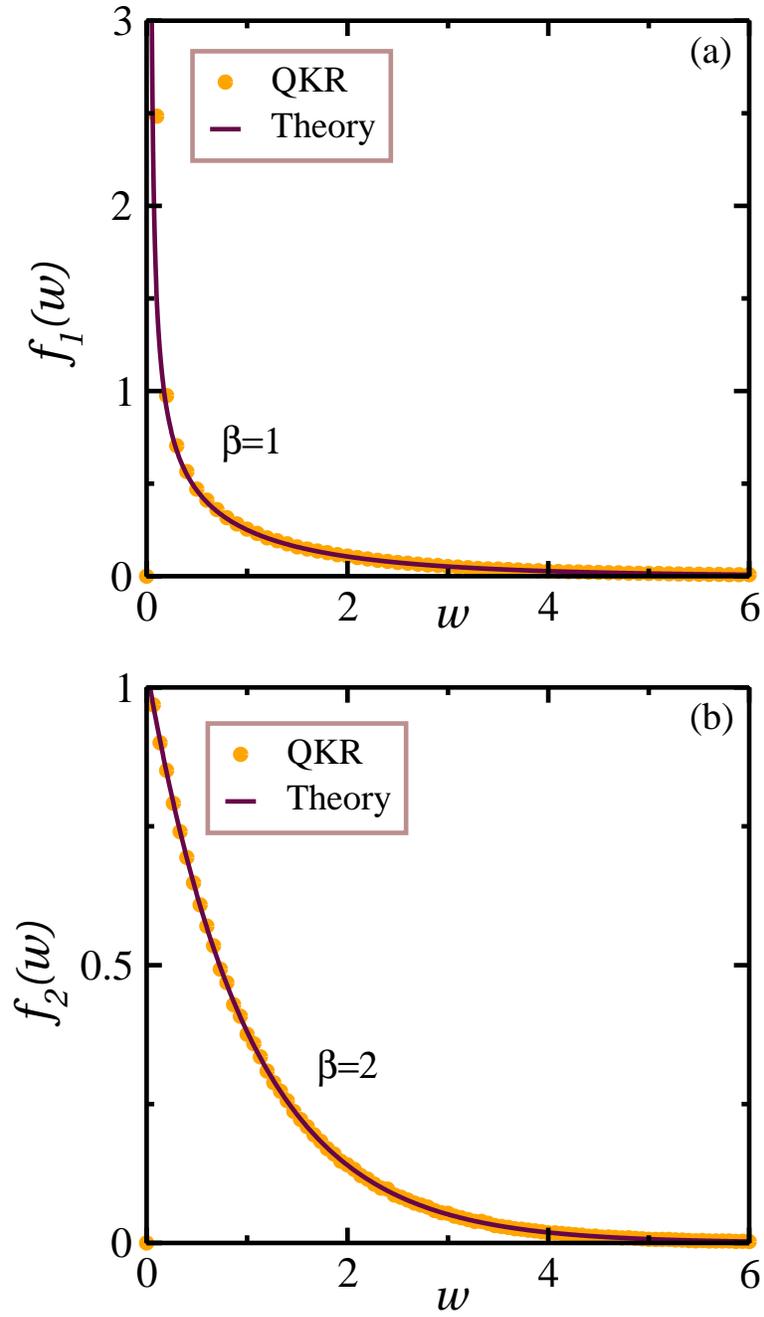}
\caption{Distribution of eigenvector components $w=N|u_{j}|^{2}$ of the QKR. We show numerical results for (a) $\gamma=0.0~(\beta=1)$, and (b) $\gamma=0.7~(\beta=2)$. The RMT results are plotted from Eq.~(\ref{Eq.36}).}\label{fig:QKR-EIGENVEC}
\end{center}
\end{figure}

In Fig.~\ref{fig:QKR-NNSD}, we plot $p_0(s)$ vs. $s$ for the QKR, and compare it with corresponding RMT results. As mentioned
earlier, $\gamma = 0.0$ yields the $\beta = 1$ case, and $\gamma = 0.7$ gives the $\beta = 2$ case. We see an excellent agreement between the QKR results and RMT. Figures~\ref{fig:QKR-NV} and \ref{fig:QKR-R2} are analogous plots of number variance and the two-point correlation function.
It is also relevant to study the statistics of eigenvectors obtained from the QKR ensembles. In Fig.~\ref{fig:QKR-EIGENVEC}, we plot $f_\beta(w)$ vs. $w$ for the QKR with $\gamma = 0.0~(\beta = 1)$ and $\gamma = 0.7~(\beta = 2)$. Again, the QKR data is in excellent agreement with the RMT results in Eq.~(\ref{Eq.36}).

\begin{figure}[H]
\begin{center}
\includegraphics[width=\linewidth, width=10cm]{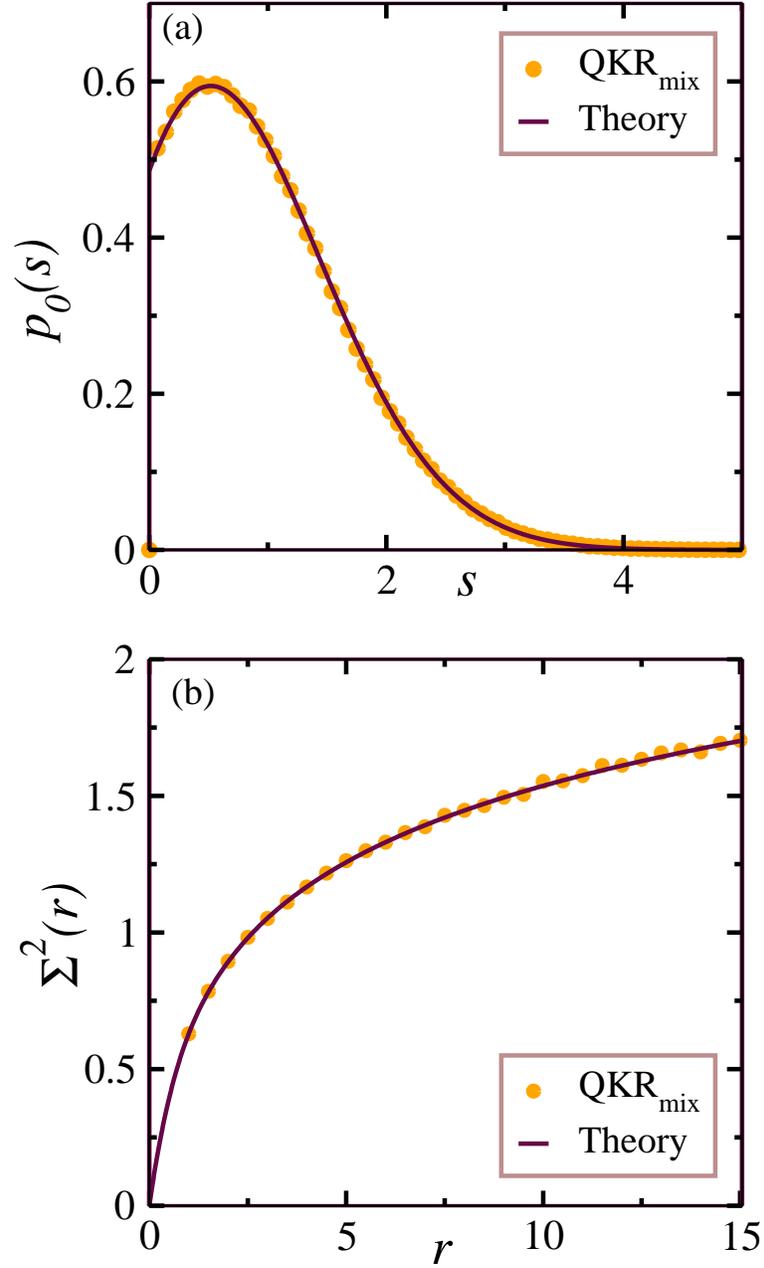}
\caption{Realization of mixed spectra from COE in QKR. The parameter values are $\gamma=0.0$ and $\phi_{0}=0$, which yields a superposition of 2 independent COE spectra (of opposite parity). We show results for (a) Nearest-neighbor spacing distribution, and (b) Number variance. The RMT result is provided in Eqs.~(\ref{Eq.32})-(\ref{Eq.33}).}\label{fig:MIXING}
\end{center}
\end{figure}

\begin{figure}[H]
\begin{center}
\includegraphics[width=\linewidth, width=14cm]{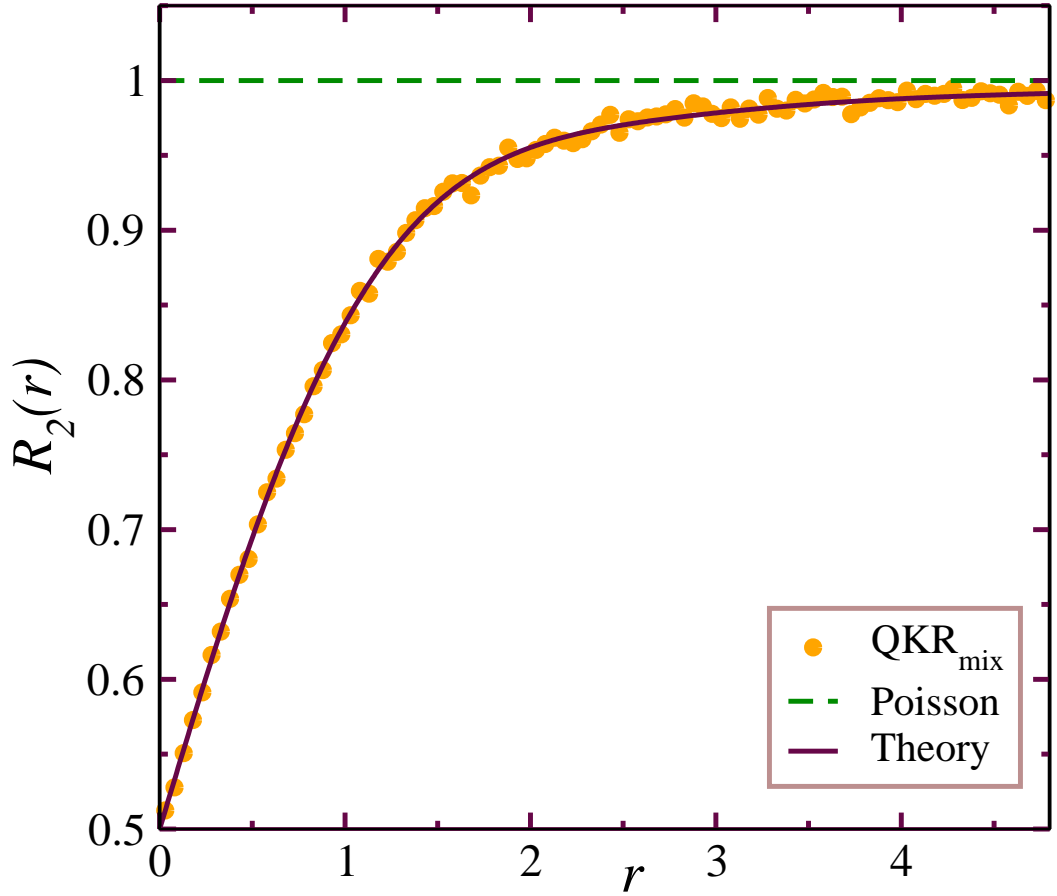}
\caption{Analogous to Fig.~\ref{fig:MIXING}, but for the two-point correlation function. The RMT result is provided in Eq.~(\ref{Eq.33c}).}\label{fig:MIXING1}
\end{center}
\end{figure}

Next, let us examine the spectra of mixed ensembles which were introduced in Eq.~(\ref{Eq.28}). We obtain mixed spectra from QKR by choosing $\gamma = 0.0$, $\phi_0=0$. The parameter value $\phi_0=0$ corresponds to the sum of 2 independent spectra arising from states of even and odd parity. Clearly, the mixed spectrum is also characterized by TRI. In Fig.~\ref{fig:MIXING}, we show results for $p_0(s)$ and $\Sigma^2(r)$ for the mixed spectrum. The QKR data is in excellent agreement with the RMT result in Eqs.~(\ref{Eq.32})-(\ref{Eq.33}). In Fig.~\ref{fig:MIXING1}, we show the two-point correlation function.
The QKR does not exhibit CSE (or GSE) spectra directly. The system of kicked tops \cite{FH} has been investigated for direct realization of all the three random matrix ensembles (COE, CUE, CSE). We discuss here a method of indirect realization of CSE in QKR. 

There are two remarkable theorems which relate the fluctuations of the three classical ensembles. The first theorem \cite{MLD} states that the spectra of CSE can be obtained by choosing alternate eigenvalues of COE. The second theorem \cite{FJDY3, GUNSON} states that the spectra of CUE can be obtained by choosing alternate eigenvalues from a random superposition of two independent COE spectra of the same dimension.

\begin{figure}[H]
\begin{center}
\includegraphics[width=\linewidth, width=10cm]{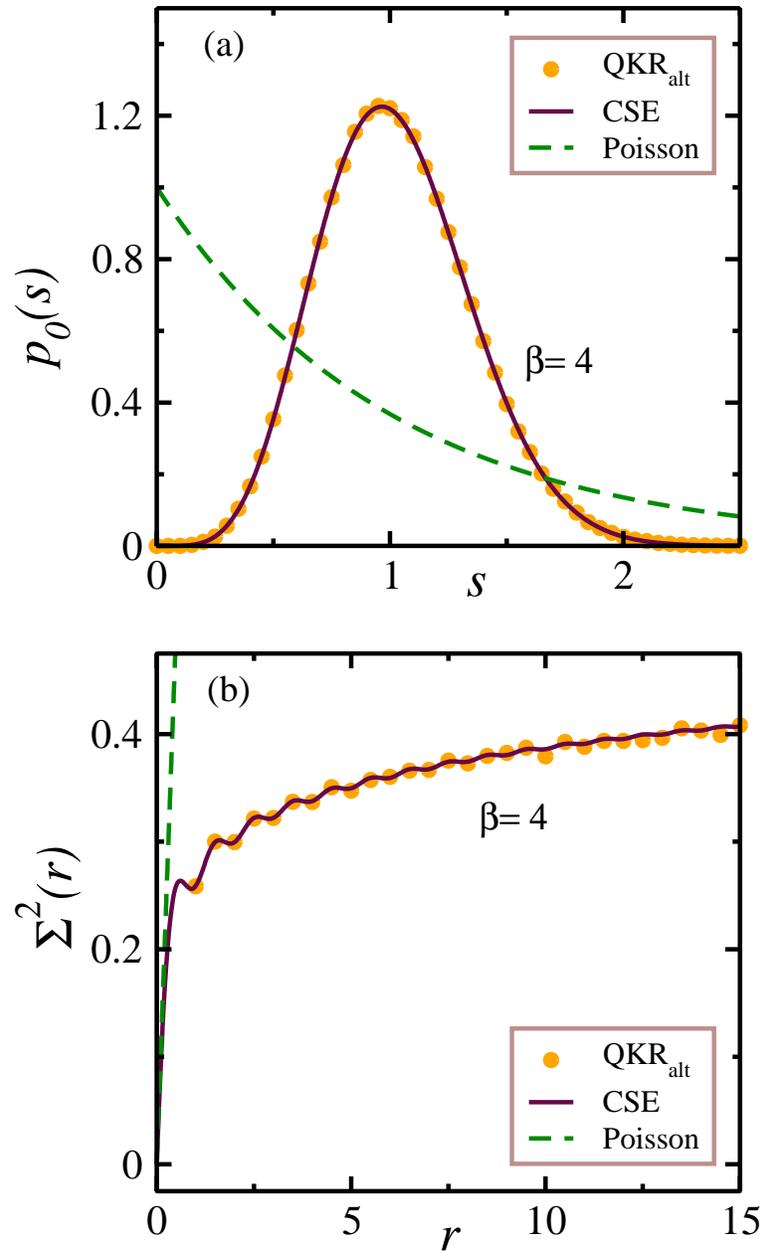}
\caption{Indirect realization of CSE ($\beta=4$) in QKR. Alternate eigenvalues of COE ($\beta=1$) spectra are analyzed giving rise to CSE statistics. The frames show (a) Nearest-neighbor spacing distribution, and (b) Number variance.}\label{fig:QKR-CSE}
\end{center}
\end{figure}

\begin{figure}[H]
\begin{center}
\includegraphics[width=\linewidth, width=14cm]{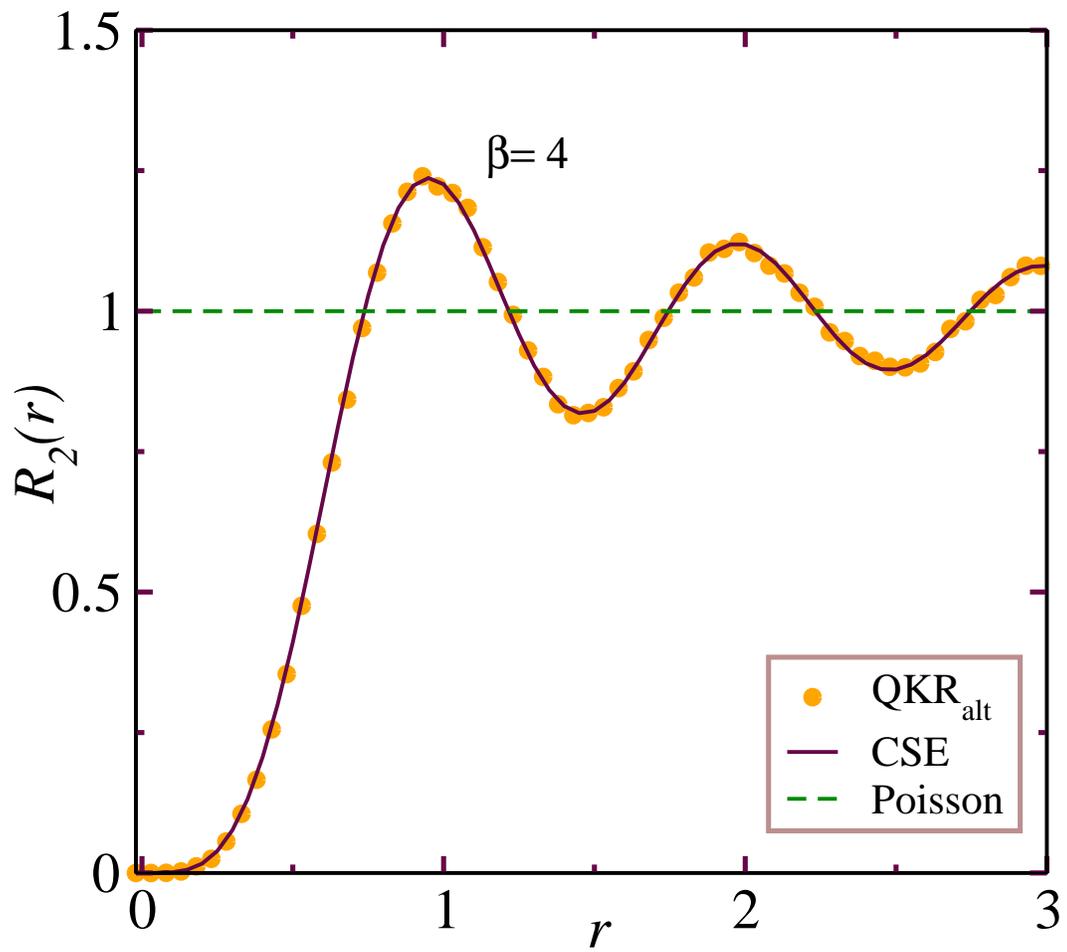}
\caption{Two-point correlation function for CSE ($\beta=4$), obtained indirectly in QKR.}\label{fig:QKR-CSE-R2}
\end{center}
\end{figure}

\begin{figure}[H]
\begin{center}
\includegraphics[width=\linewidth, width=10cm]{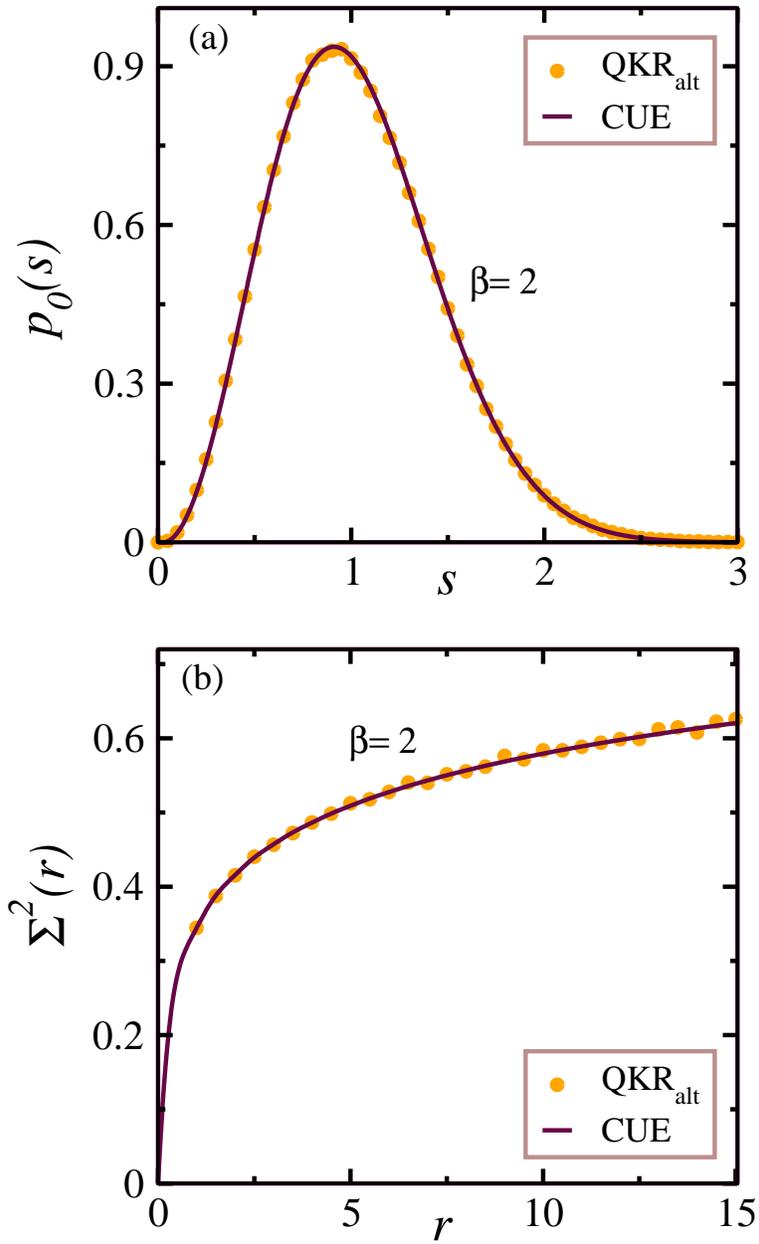}
\caption{Indirect realization of CUE ($\beta=2$) from COE in QKR. In this case, $\gamma=0.0$ and $\phi_{0}=0$, which preserves both the TRI and parity, giving rise to a superposition of two independent COE spectra. The analysis of alternate eigenvalues of this spectrum gives CUE statistics. We show data for (a) Nearest-neighbor spacing distribution, and (b) Number variance.} \label{fig:MIXING-ALT}
\end{center}
\end{figure}

\begin{figure}[H]
\begin{center}
\includegraphics[width=\linewidth, width=14cm]{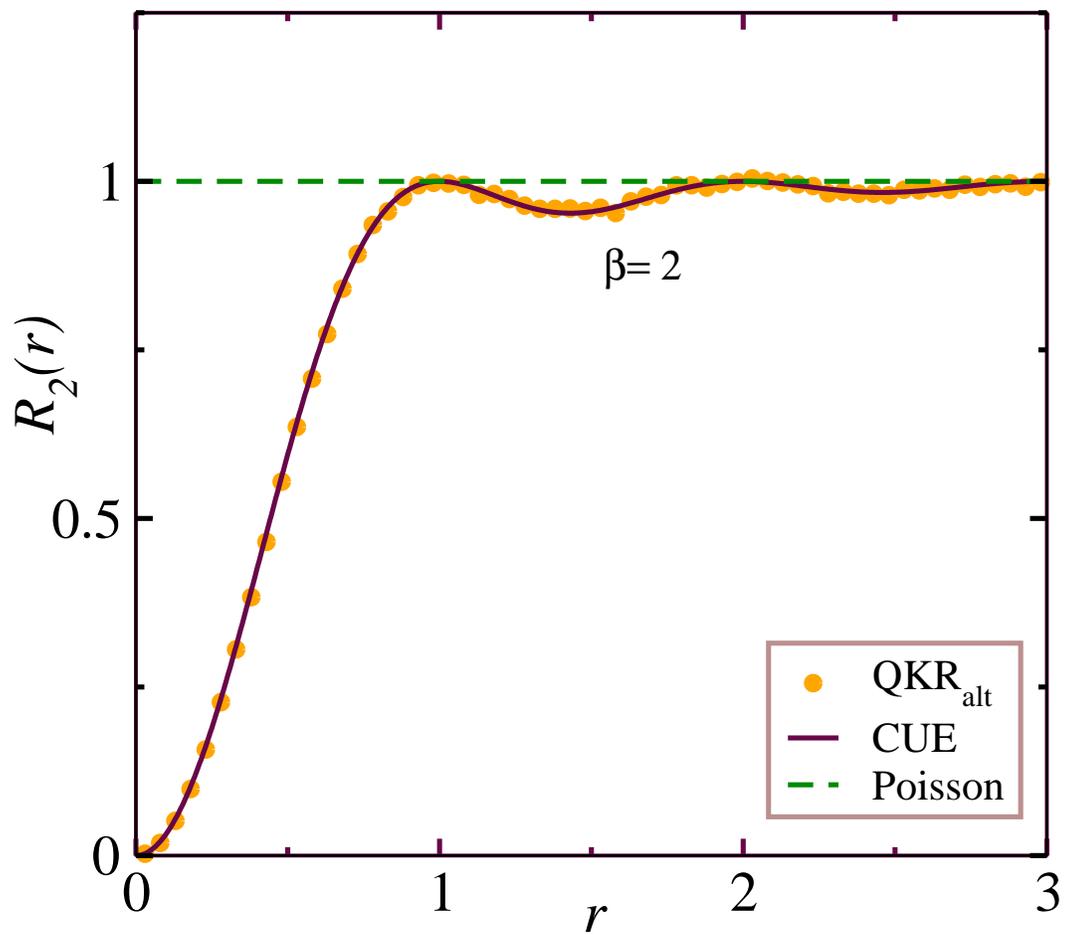}
\caption{Analogous to Fig.~\ref{fig:MIXING-ALT}, but for the two-point correlation function.} \label{fig:MIXING-ALT1}
\end{center}
\end{figure}

In Fig.~\ref{fig:QKR-CSE}, we show the spacing distribution and the number variance of alternate eigenvalues of the QKR ensemble (of size $1000$) with matrix dimension $N=1000$. The parameter values are $\gamma=0.0$ and $\phi_{0}=\pi/(2N)$. In this case, the dimension of the matrix reduces to $500$ and the number of the spectra becomes $2000$. The results show excellent agreement with CSE. In Fig.~\ref{fig:QKR-CSE-R2}, we show the analogous plot for the two-point correlation function.
In a similar fashion, we consider QKR spectra for $\gamma=0.0, \phi_{0}=0$. As mentioned earlier, since $\phi_{0}=0$, this leads to a superposition of two independent COE spectra with opposite parities. An analysis of alternate eigenvalues gives excellent agreement with CUE, as seen in Figs.~\ref{fig:MIXING-ALT} and \ref{fig:MIXING-ALT1}.

\section{Transition Ensembles}\label{S11}

The ensembles introduced in Sec.~\ref{S2} correspond to exact symmetries, i.e., a symmetry is either fully preserved or fully broken. In this section, we briefly consider ensembles which have a partially broken symmetry. Let us consider a Gaussian ensemble defined by \cite{FKPTV}
\begin{equation}\label{Eq.51}
H_{\alpha}= H(S) + i\alpha H(A).
\end{equation}
Here $H(S)$ are real symmetric matrices, $H(A)$ are real antisymmetric matrices, and $\alpha$ is a real parameter. The distinct matrix elements of $H(S)$ and $H(A)$ are independent Gaussian random variables with zero mean. The variances are $v^{2}$ for the off-diagonal elements of $H(S)$ and $H(A)$, and $2v^{2}$ for the diagonal matrix elements of $H(S)$. The diagonal matrix elements of $H(A)$ are zero. Note that the $H_{\alpha}$-ensemble is GOE for $\alpha=0$, and GUE for $\alpha=1$. Thus $\alpha$ is a measure of TRI breaking in the system and parametrizes the GOE $\rightarrow$ GUE transition. 

This problem has been solved approximately in \cite{FKPTV}, and exactly in \cite{PM}. The average density of eigenvalues is again the semicircle of Eq.~(\ref{Eq.5}) with
\begin{equation}\label{Eq.52}
R^{2}= 4Nv^{2}(1+\alpha^{2}).
\end{equation}
For large $N$, the exact two-level cluster function is given by \cite{PM}
\begin{eqnarray}\label{Eq.53}
Y_{2}(r, \Lambda)&=&\left( \frac{\sin \pi r}{\pi r}\right)^{2}-\frac{1}{\pi^{2}}\int_{0}^{\pi}dx~x~\sin(xr)\exp(2\Lambda x^{2})\times \nonumber \\
&&~~~~~~~~~~~~\int_{\pi}^{\infty} dy~\frac{\sin(yr)}{y}\exp(-2\Lambda y^{2}),
\end{eqnarray}
where $\Lambda$ is the transition parameter:
\begin{equation}\label{Eq.54}
\Lambda = \alpha^{2}v^{2}/(D(x))^{2}.
\end{equation}
Here $D=1/[N \bar{\rho}(x)]$ is the average spacing at $x$. For $\Lambda=0, \infty$, we obtain the GOE and GUE results of Eqs.~(\ref{Eq.11}), (\ref{Eq.12}) respectively. Note that, for spectra with finite span, one should choose $v^{2}N=1$ and therefore $\Lambda\propto \alpha^{2}N$. Thus, as $N$ increases, $\alpha$ becomes smaller to keep $\Lambda$ finite. For $N=\infty$, the GOE$\rightarrow$ GUE transition is abrupt at $\alpha=0$. 

The exact spacing distributions are also known for this problem \cite{MP}. As in Sec.~\ref{S3}, these are complicated but the nearest-neighbor spacing distribution for two-dimensional matrices gives an excellent approximation to the $N\rightarrow \infty$ result. The $N=2$ result is \cite{FKPTV}
\begin{equation}\label{Eq.55}
p_{0}(s)=\frac{s}{4v^{2}(1-{\alpha^\prime}^{2})^{1/2}}\exp\left(  -\frac{s^2}{8v^2}\right) \text{erf}\left[ \left( \frac{1-{\alpha^\prime}^{2}}{8{\alpha^\prime}^{2}v^{2}}\right) s\right] .
\end{equation}
Here, instead of Eq.~(\ref{Eq.54}), one has to use $\alpha^{\prime}$ as a fitting parameter to obtain the spacing distribution.

In actual applications, $\Lambda$ should be interpreted as the symmetry-breaking matrix elements of the relevant Hamiltonian. These results also apply to circular ensembles with the Hamiltonian being replaced by the unitary operator \cite{PS}.

The number variance, $\Sigma^{2}(r;\Lambda)$, can be obtained by numerical integration of Eq.~(\ref{Eq.15}) with $Y_{2}$ from Eq.~(\ref{Eq.53}). However, a very good approximation can be written as 
\begin{equation}\label{Eq.56}
\Sigma^{2}(r;\Lambda)= \Sigma_{\beta=2}^{2}(r)+\frac{1}{2\pi^2}\ln\left[ 1+\frac{\pi^{2}r^{2}}{4(\tau+2\pi^{2}\Lambda)^{2}}\right] ,
\end{equation}
with $\tau=0.615$.

Let us consider some important applications of the GOE $\rightarrow$ GUE transition. This transition has proved to be useful in deriving upper bounds on the TRI breaking part of the nucleon-nucleon interaction \cite{FKPTV, FKPTVI}. Another system where this transition is applicable is the Aharonov-Bohm chaotic billiard, where a single line of magnetic flux passes perpendicular through the plane of the system, breaking TRI \cite{BR}. A calculation involving Aharonov-Bohm flux lines in a disordered metallic ring again shows the exact GOE $\rightarrow$ GUE realization of $\Sigma^{2}(r)$ and $p_{0}(s)$ \cite{DM}.

The parameter $\Lambda$ also appears in transitions involving breaking of other symmetries. Then, one considers the generalization of Eq.~(\ref{Eq.51}) as
\begin{equation}\label{Eq.57}
H(\alpha)= A+\alpha B,
\end{equation}
where $A$ and $B$ are independent Gaussian random matrices representing symmetry-preserving and symmetry-breaking parts of the Hamiltonian respectively. In such cases, $v^{2}$ in Eq.~(\ref{Eq.54}) refers to the variance of the symmetry-breaking matrix elements $B_{jk}$ in the $A$-diagonal representation.

An example is the GSE $\rightarrow$ GUE transition, which has also been solved exactly \cite{MP}. In this case, the exact two-level cluster function is given by
\begin{eqnarray}\label{Eq.58}
Y_{2}(r, \Lambda)&=&\left( \frac{\sin \pi r}{\pi r}\right)^{2}-\frac{1}{\pi^{2}}\int_{0}^{\pi}dx~\frac{\sin(xr)}{x}\exp(2\Lambda x^{2})\times \nonumber \\
&&~~~~~~~~~~~~\int_{\pi}^{\infty} dy~y\sin(yr)\exp(-2\Lambda y^{2}).
\end{eqnarray}
For $\Lambda=0$, Eq.~(\ref{Eq.58}) reduces to a modified version of Eq.~(\ref{Eq.13}), where the double degeneracy of the eigenvalues is explicitly taken into account. Eqs.~(\ref{Eq.53}) and (\ref{Eq.58}) also apply to transitions in Jacobi ensembles \cite{KP1}.

For studies of parity breaking, one considers the 2GOE $\rightarrow$ 1GOE transition. More generally, one can consider $l$GOE$ \rightarrow$ $1$GOE transitions for the breaking of symmetries having $l$ quantum numbers. Here, $l$GOE refers to a direct sum of $l$ independent GOEs. For large $l$, this becomes the Poisson $\rightarrow$ GOE transition. An example of this has been found in complex atomic spectra \cite{RP}, where the Poisson $\rightarrow$ GOE transition occurs because of the breaking of the LS symmetry.

\begin{figure}[H]
\begin{center}
\includegraphics[width=\linewidth, width=14cm]{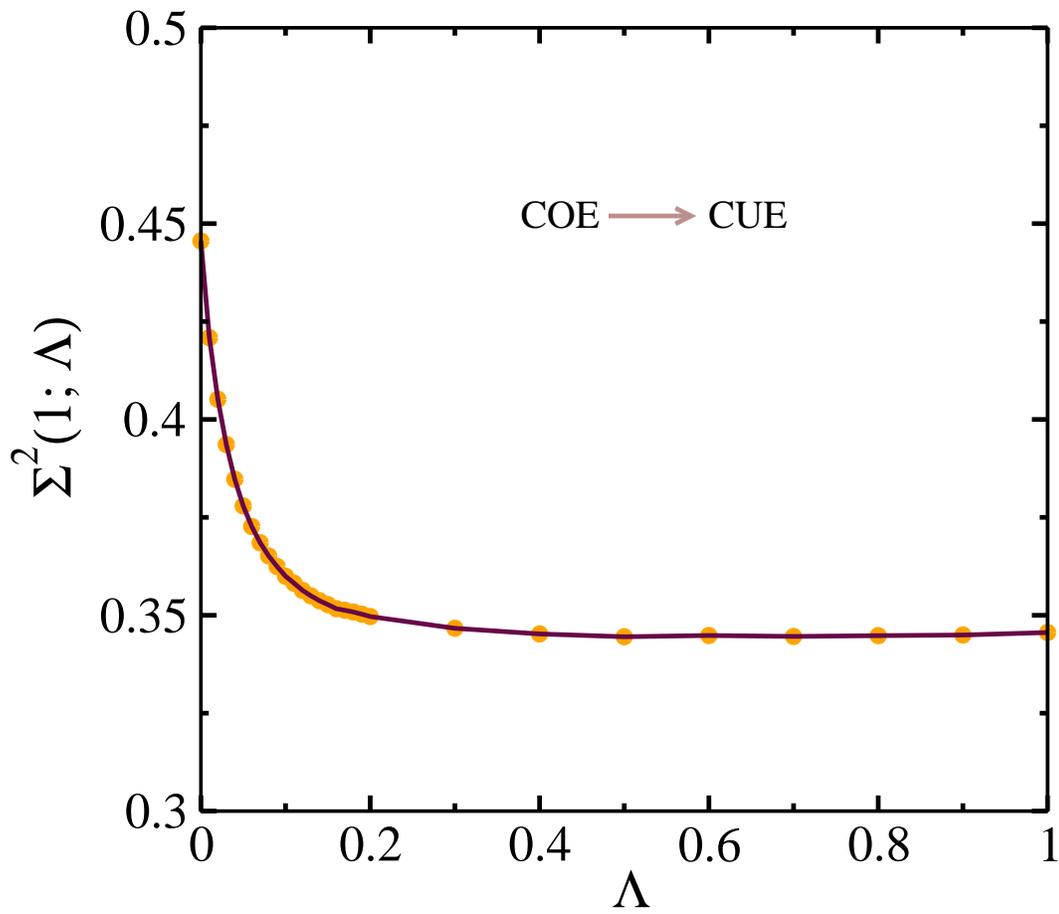}
\caption{Number variance $\Sigma^{2}(r;\Lambda)$ vs. $\Lambda$ for $r=1$, showing the transition from COE $(\beta=1)$ $\rightarrow$ CUE $(\beta=2)$.}\label{fig:TRANSITION}
\end{center}
\end{figure}

\begin{figure}[H]
\begin{center}
\includegraphics[width=\linewidth, width=9cm]{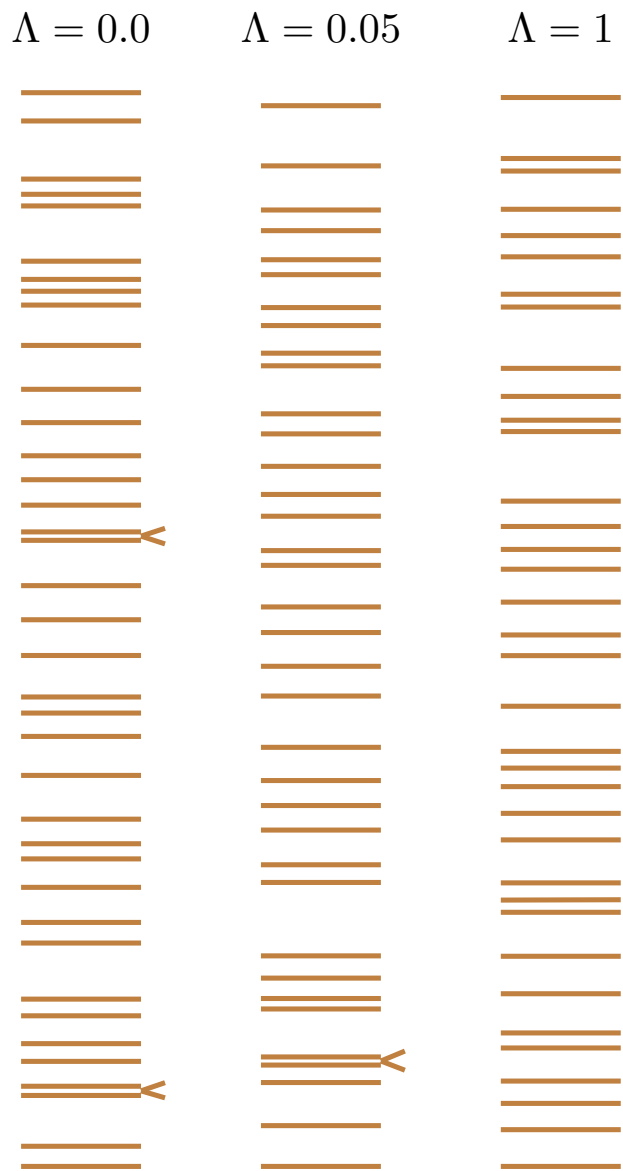}
\caption{Spectral levels of COE $\rightarrow$ CUE transition ensemble for $\Lambda = 0.0, 0.05, 1.0$. The arrowheads mark the occurrences of pairs of levels with spacings smaller than $1/2$ of the average spacing.}\label{fig:INTERMEDIATE-TRANSITION}
\end{center}
\end{figure}

As mentioned earlier, the Gaussian transition results are obtained for transitions in circular ensembles also. (For other transitions, see \cite{APF}.) Let us illustrate the COE $\rightarrow$ CUE transition using the QKR. In QKR, $\gamma p$ plays the role of the TRI breaking operator, with $p$ being the momentum operator. For large $N$, the trace of the matrix $\gamma p$ is $(\gamma^{2}N^{3})/12$, and the mean square of the matrix elements $\left( \overline{|(\gamma p)_{jk}|^{2}}\right)$ is $\gamma^{2}N/12$. Using $D(x)=2\pi/N$, $\Lambda$ in the QKR is given by
\begin{equation}\label{Eq.59}
\Lambda= \frac{ \gamma^{2}N^{3} }{48\pi^{2}}.
\end{equation}
In Fig.~\ref{fig:TRANSITION}, we plot $\Sigma^{2}(1,\Lambda)$ vs. $\Lambda$. (See also \cite{PRS} for a demonstration of this transition.) In Fig.~\ref{fig:INTERMEDIATE-TRANSITION}, we show the change in the spectrum for the COE $\rightarrow$ CUE transition with $\Lambda=0.0, 0.05, 1.0$.

In QKR, one can also find 2COE $\rightarrow$ 1COE by varying $\phi_{0}$ and keeping $\gamma=0$ \cite{PRS}. Here, 2COE corresponds to a direct sum of two independent COEs, as discussed earlier. This kind of transition can be useful in studying parity-breaking in real systems. 

Apart from the spectral transitions, we can also study the transitions in the eigenvectors \cite{SP}. We will not discuss this subject further here.

\section{Conductance fluctuations in Mesoscopic Systems}\label{S12}

Mesoscopic physics deals with systems which are intermediate in size between the atomic scale and the macroscopic scale. The
transport properties of metals and insulators at very small scales have given new insights in understanding mesoscopic physics. The quantities of special interest in this context are the conductance fluctuations and distributions. In this section, we briefly discuss the application of RMT to understand transport properties of mesoscopic systems. There has been extensive study of these systems in the literature, and we refer the interested reader to some important review articles \cite{CWB, GMW, ALHS, AM, KP1}.

The most important phenomenon in this context is that of `universal conductance fluctuations' (UCF) in quantum dots. This refers to the independence of conductance fluctuations (measured in units of $e^2/\hbar$) from the sample size, degree of disorder, and other parameters of the system.

\begin{figure}[H]
\begin{center}
\includegraphics[width=\linewidth, width=16cm]{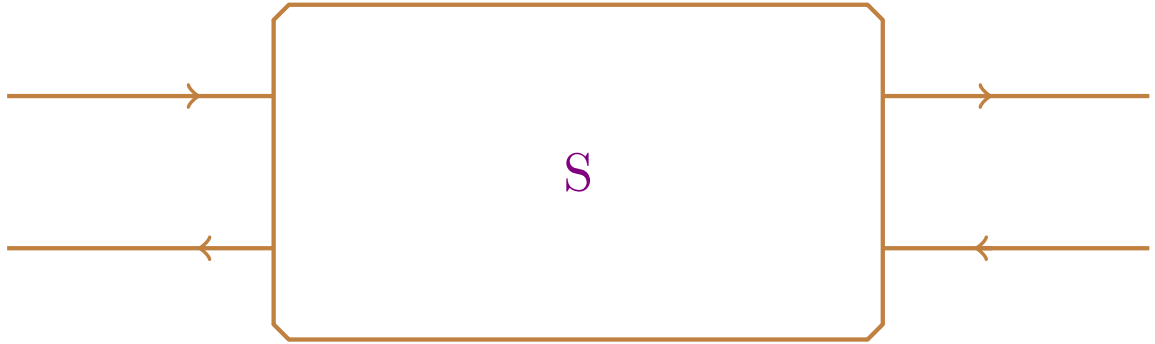}
\caption{Schematic of the scattering matrix for a quantum dot.}\label{fig:QUANTUM-DOT}
\end{center}
\end{figure}

The scattering matrix $S$ for the conductance problem, shown schematically in Fig.~(\ref{fig:QUANTUM-DOT}), has the standard decomposition in terms of reflection matrices $r, r^{\prime}$ and transmission matrices $t, t^{\prime}$ as
\begin{equation}\label{Eq.60}
S=
\begin{pmatrix}
r_{N_{1}\times N_{1}} & t_{N_{1}\times N_{2}}^{\prime} \\
t_{N_{2}\times N_{1}} & r_{N_{2}\times N_{2}}^{\prime}
\end{pmatrix}.
\end{equation}
Here, $N_{1}$ and $N_{2}$ refer to the number of modes in the left and right leads connected to the cavity. The matrices $r_{N_{1}\times N_{1}}$ and $r_{N_{2}\times N_{2}}^{\prime}$ correspond to the reflection from left-to-left and right-to-right, respectively. Similarly, $t_{N_{2}\times N_{1}}$ and $t_{N_{1}\times N_{2}}^{\prime}$ represent the transmission from left-to-right and right-to-left, respectively. The scattering matrix $S$ is thus $N_{s}$-dimensional, where $N_{s}= N_{1}+N_{2}$. We also define $N=$min$(N_{1},N_{2})$. As a consequence of unitarity of $S$, the Hermitian matrices $t^{\dagger}t$, ${t^{\prime}}^{\dagger}t^{\prime}$, $1-r^{\dagger}r$ and  $1-{r^{\prime}}^{\dagger}r^{\prime}$ have $N$ common eigenvalues $T_{1},\cdots, T_{N}$ with values between $0$ and $1$ \cite{KP3}. We denote the set of these transmission eigenvalues as $\{T_i\}$.

Using the circular ensembles for $S$, it can be shown that the jpd of transmission eigenvalues $p(\{T_i\})$ has a form analogous to Eqs.~(\ref{Eq.2})-(\ref{Eq.4}) with 
\begin{equation}\label{Eq.61}
p(T_{1},\cdots,T_{N})= C^{\prime \prime}\prod_{j<k}|T_j-T_k|^{\beta} \prod_{j=1}^{N} T_j^{\frac{\beta}{2}(|N_{1}-N_{2}|+1-\frac{2}{\beta})},
\end{equation}
and
\begin{equation}\label{Eq.62}
V(T)= -\frac{1}{2}\left( \vert N_{1}-N_{2}\vert +1 -\frac{2}{\beta} \right) \ln T.
\end{equation}

The dimensionless conductance at zero temperature is related to the eigenvalues by the Landauer formula \cite{ALT, LS, LSF, LANDAU1, LANDAU2, IMRY1, IMRY2, MPS, SMMP}
\begin{equation}\label{Eq.63}
g=\sum_{j=1}^{N}T_{j}.
\end{equation}
The average and variance of $g$ can be calculated from the correlation functions of the jpd in Eq.~(\ref{Eq.61}). For $N_{1}, N_{2} \gg 1$, these are given by
\begin{equation}\label{Eq.64}
\bar{g}= \frac{N_{1}N_{2}}{N_{s}} {}{}{}{} \xrightarrow[]{N_{1}=N_{2}} {} \frac{N}{2},
\end{equation}
and
\begin{equation}\label{Eq.65}
\text{var}(g)= \frac{2N_{1}^{2}N_{2}^{2}}{\beta N_{s}^{4}}  {}{}{}{} \xrightarrow[]{N_{1}=N_{2}} {} \frac{1}{8 \beta}.
\end{equation}
Equation~(\ref{Eq.65}) for the variance of $g$ describes the UCF.

The dimensionless shot-noise power is given by the B\"{u}ttiker formula:
\begin{equation}\label{Eq.66}
p= \sum_{j=1}^{N} T_{j}(1-T_{j}).
\end{equation}
Its average and variance for large $N_{1}$ $N_{2}$ are given by 
\begin{equation}\label{Eq.67}
\bar{p}= \frac{N_{1}^{2}N_{2}^{2}}{N_{s}^{3}} {}{}{}{} \xrightarrow[]{N_{1}=N_{2}} \frac{N}{8} ,
\end{equation}
\begin{equation}\label{Eq.68}
\text{var}(p)= \frac{4N_{1}^{4}N_{2}^{4}}{\beta N_{s}^{8}} {}{}{}{} \xrightarrow[]{N_{1}=N_{2}} {}{}{}\frac{1}{64 \beta}.
\end{equation}

\begin{figure}[H]
\begin{center}
\includegraphics[width=\linewidth, width=16cm]{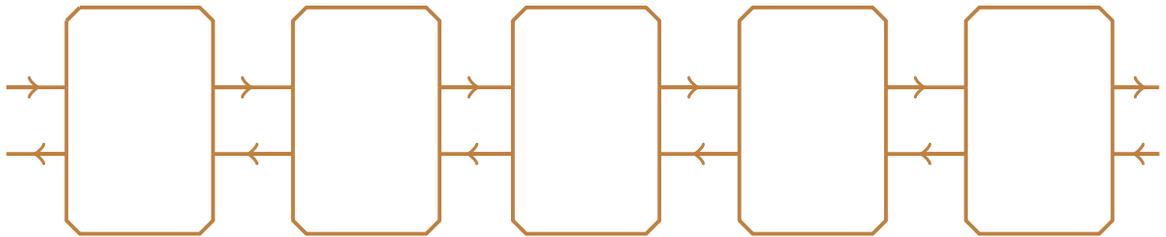}
\caption{Schematic of a disordered nanowire. Each segment is analogous to a quantum dot.}\label{fig:WIRE}
\end{center}
\end{figure}

As mentioned earlier, the above results apply for quantum dots. There have also been many studies of disordered nanowires, which may be modeled as a sequence of coupled quantum dots (see Fig.~\ref{fig:WIRE}). This system was first studied in a different theoretical framework (independent of RMT) using diagrammatic perturbation theory \cite{ALT}.

The corresponding result for conductance fluctuations is 
\begin{equation}\label{Eq.69}
\text{var}(g)= \frac{2}{15 \beta} ,
\end{equation}
which differs slightly from the RMT result for quantum dots. Subsequently, a Brownian motion model was developed using RMT to re-derive Eq.~(\ref{Eq.69}). The corresponding diffusion equation is referred to as the Dorokhov-Mello-Pereyra-Kumar (DMPK) equation \cite{CWB, CWBHB}.

\section{Finite Range Coulomb Gas Models}\label{S13}

In Eqs.~(\ref{Eq.2})-(\ref{Eq.4}), all particles have pairwise interactions. In this section, we consider the natural generalization to the case where particles have finite-range interactions. We refer to these ensembles as finite-range Coulomb gas (FRCG) models \cite{PKP}. For linear ensembles, the jpd is described by Eqs.~(\ref{Eq.2})-(\ref{Eq.4}) with $|j-k|\leq d$, where $d$ is the range of the interaction. A similar definition applies for circular ensembles with $x_j \rightarrow e^{i \theta_j}$.

For $N\gg d$, the linear and circular ensembles again give identical fluctuation results. The fluctuation properties are characterized by $d$. It has been shown that the FRCG models are exactly solvable for each $d$ \cite{PKP}. We present here some analytic results, supplemented by MC results.

For $d=0$, there is no interaction between particles, and we find the Poisson results:
\begin{eqnarray}
p_{n-1}(s) &=& \frac{s^{n-1}}{(n-1)!}e^{-s}, \label{Eq.72} \\
\Sigma^{2}(r) &=& r. \label{Eq.73}
\end{eqnarray}
For $d=1$, we obtain
\begin{eqnarray}
p_{n-1}(s) &=& \frac{(\beta+1)^{n(\beta+1)}}{\Gamma(n(\beta+1))}s^{n(\beta+1)-1}e^{-(\beta+1)s} , \label{Eq.74} \\
\Sigma^{2}(r) &=& \frac{r}{\beta+1}+\frac{\beta(\beta+2)}{6(\beta+1)^{2}}. \label{Eq.75}
\end{eqnarray}

For $d\geq 2$, analytic results can be derived from an integral equation approach \cite{PKP}. However, for large $k$, there is a simple mean-field (MF) approximation, which reduces the arbitrary-$d$ problem to an effective $d=1$ problem. The MF results are
\begin{eqnarray}
p_{n-1}(s) &=& \left[\Gamma(n\xi)\right]^{-1} \xi^{n\xi} s^{n\xi -1}e^{-\xi s}, \label{Eq.76} \\
\Sigma^{2}(r) &=& \frac{r}{\xi} + \frac{(\xi^2 -1)}{6\xi^{2}} , \label{Eq.77}
\end{eqnarray}
where $\xi=\beta d +1$.

We have found that the FRCG models describe the spectra of QKR with the identification $d=\alpha^{2}/N$ \cite{PKP}. (The importance of the parameter $\alpha^{2}/N$ was earlier emphasized by Casati and others, who established an empirical relationship between QKR and banded random matrices.) As $ \alpha$ is a continuous parameter, this allows $d$ to take non-integer values also. We have formulated FRCG models for fractional $d$ \cite{PKP}, which we do not discuss here.

\begin{figure}[H]
\begin{center}
\includegraphics[width=\linewidth, width=8cm]{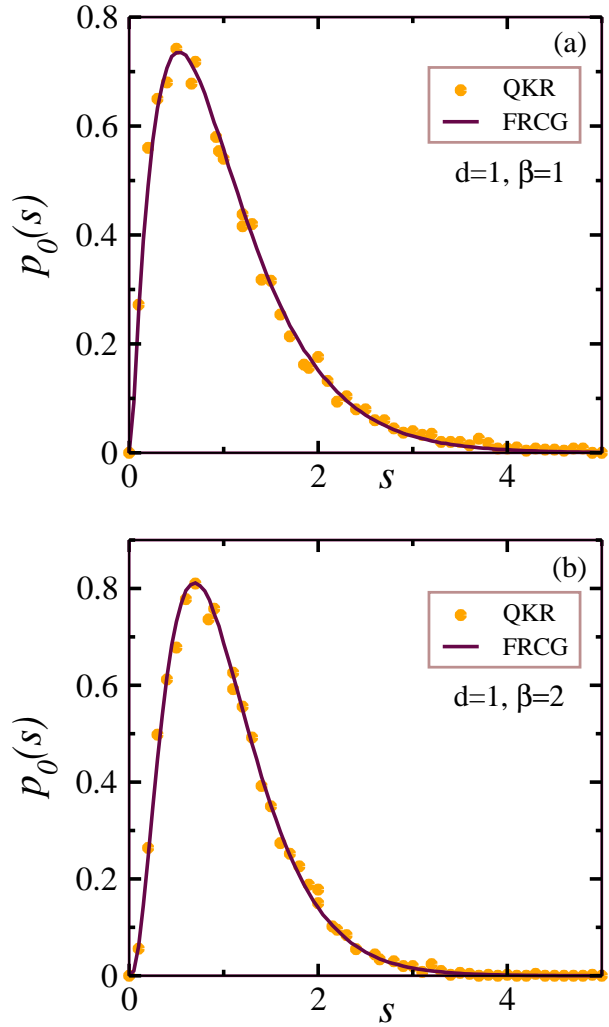}
\caption{Nearest-neighbor spacing distribution [$p_0(s)$ vs. $s$] for $ d=1$, and (a) $\beta=1$, (b) $\beta=2$. The QKR data is obtained for $\alpha = \sqrt{dN}$. The FRCG result is given in Eq.~(\ref{Eq.74}).}\label{fig:FRCG1}
\end{center}
\end{figure}

\begin{figure}[H]
\begin{center}
\includegraphics[width=\linewidth, width=12cm]{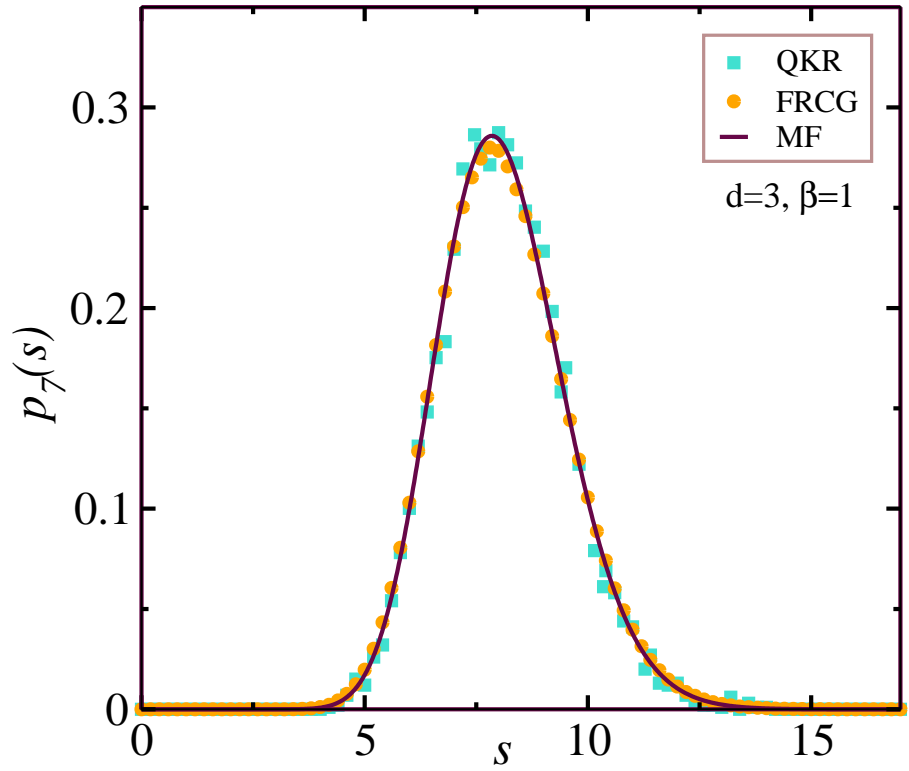}
\caption{Plot of $p_7(s)$ vs. $s$ for the QKR and FRCG with $d=3, \beta=1$. The FRCG result is obtained via MC. For purposes of comparison, we also show the MF result in Eq.~(\ref{Eq.76}).}\label{fig:FRCG2}
\end{center}
\end{figure}

In Fig.~\ref{fig:FRCG1}, we plot $p_{0}(s)$ vs. $s$ for the QKR with $\beta=1,2$. The parameter $\alpha = \sqrt{dN}$ with $d=1$. We also plot the corresponding FRCG result from Eq.~(\ref{Eq.74}). The agreement between QKR and FRCG results is excellent. In Fig.~\ref{fig:FRCG2}, we show results for $p_{7}(s)$  vs. $s$ for $d=3,\beta=1$. Here QKR, FRCG and MF results are compared. Again, the agreement is very good.

\section{Wishart Ensembles}\label{S14}

As mentioned in Sec.~\ref{S1}, random matrices were first introduced by Wishart in the study of statistical multivariate analysis. Subsequent interest in RMT focused on physics applications. Recently, there has been a resurgence of interest in Wishart ensembles and their generalizations. These have found applications in multivariate analysis of time series which arise in, e.g., chaotic physical systems \cite{VAP}, economics \cite{EC1, EC2}, disordered solids \cite{SMP}, biological systems \cite{BIO1, BIO2, BIO3, BIO4}, meteorology \cite{ATOM1, ATOM2, ATOM3}, and communication theory \cite{COM1, COM2, COM3}.

In this section, we briefly review Wishart ensembles and our theoretical understanding of them. We consider matrices of the form
\begin{equation}\label{Eq.78}
H= AA^{\dagger},
\end{equation}
where the elements of $A$ can be real ($\beta=1$), complex ($\beta=2$), or quaternion ($\beta=4$). Each of the $\beta$ components of the matrix elements of $A$ is an independent and identically distributed (i.i.d.) Gaussian random variable with variance $1/2$. For  example, in an economics application, each row of $A$ may consist of a time series for a particular stock price. The different rows correspond to different stocks.

In general, $A$ is an $N \times M$ rectangular matrix, where $N$ denotes the number of stocks, and $M$ denotes the number of evenly-spaced times. The corresponding $H$ will be an $N \times N$ matrix. The jpd of eigenvalues of $H$ has the form in Eq.~(\ref{Eq.4}):
\begin{equation}\label{Eq.79}
p(x_{1},\cdots, x_{N})= C^{\prime\prime\prime}\prod_{j<k}|x_j-x_k|^{\beta} \prod_{j=1}^{N} \left( x_{j}^{\omega}e^{x_{j}}\right) ,
\end{equation}
where
\begin{align}\label{Eq.80}
&\omega= \left( \frac{\beta}{2}\right) \left( M-N+1 \right) -1,  & N\leq M.
\end{align}

Note that all three distributions which we have mentioned in this article, viz., Gaussian ensembles with $V(x) \sim x^2$ in Eq.~(\ref{Eq.4}), ensemble of transmission eigenvalues in Eq.~(\ref{Eq.61}), and the Wishart ensemble in Eq.~(\ref{Eq.79}), are examples of general Jacobi weight functions.

\section{Conclusion}\label{S15}

In this article, we have attempted to provide a broad overview of random matrix theory (RMT) and its applications at a pedagogical and accessible level. Let us briefly summarize the topics we have covered. We have defined various types of random matrix ensembles. In physics applications, RMT applies in quantum chaotic systems, i.e., systems whose classical counterparts are chaotic.
Typically, one is interested in the properties of eigenvalues and eigenvectors of these matrix ensembles. We have discussed the various statistical measures used to quantify these properties. In particular, we have shown that the spectra of random matrices display level repulsion and spectral rigidity. Eigenvectors of random matrices also have universal properties, which are experimentally observable.

We have used the system of quantum kicked rotors (QKR) to illustrate many of the features of random matrix ensembles. Some of the most important applications of RMT have been in the context of quantum chaos in mesoscopic and nanoscopic systems. However, we should stress that present-day applications of RMT are not confined to physics alone. The framework of RMT provides insights on complexity in diverse disciplines.

\section*{Appendix A}\label{SA}

For $\beta=4$, one needs to deal with self-dual hermitian matrices. Consider a $2N\times2N$ matrix $A$:
\begin{equation}
A= B_{0}e_{0} +  B_{1}e_{1}+  B_{2}e_{2} +  B_{3}e_{3}.
\end{equation}
Here, the $B_i$ are $N$-dimensional matrices, and $ e_0,e_1,e_2, e_3$ are a two-dimensional representation of quaternions \cite{ML, CEP}:
\begin{equation}\label{Eq.81}
e_{0}=
\begin{pmatrix}
1 & 0 \\
0 & 1
\end{pmatrix},
e_{1}=
\begin{pmatrix}
0 & -i \\
-i & 0
\end{pmatrix},
e_{2}=
\begin{pmatrix}
0 & -1 \\
1 & 0
\end{pmatrix}, 
e_{3}=
\begin{pmatrix}
-i & 0 \\
0 & i
\end{pmatrix}.
\end{equation}
Thus, in quaternion space, the matrix elements of $A$ can be written as linear combination of quaternions $e_{j}$.

The dual of $A$ is defined as
\begin{equation}
A^{D}= B_{0}e_{0} -  B_{1}e_{1} - B_{2}e_{2} -  B_{3}e_{3}.
\end{equation}
The transpose of $A$ in the quaternion space is defined as 
\begin{equation}
A^{T}= B_{0}^{T}e_{0} +  B_{1}^{T}e_{1}+  B_{2}^{T}e_{2} +  B_{3}^{T}e_{3}.
\end{equation}
Here, $B^{T}$ is the transpose of $B$. $A$ is said to be self-dual if $A^{T}=A^{D}$. The matrix $A$ is Hermitian in the $(2N)$-dimensional space if $A$ is real symmetric and $B_{1}, B_{2}, B_{3}$ are real antisymmetric. For a self-dual matrix $A$, the eigenvalues are doubly degenerate.

\newpage





%%%%%%%%%%%%%%%%%%%%%%5
\end{document}